\documentclass[aps,pra,superscriptaddress,twocolumn]{revtex4-1}

\usepackage{graphicx}
\usepackage{amsmath}
\usepackage{amssymb}
\usepackage{comment}

\newcommand{\kk}{\mathbf{k}}
\newcommand{\rr}{\mathbf{r}}
\newcommand{\jj}{\mathbf{j}}
\newcommand{\J}{\mathbf{J}}
\newcommand{\A}{\mathbf{A}}

\newcommand{\pp}{\mathbf{p}}

\newcommand{\eF}{\epsilon_{\mathrm{F}}}
\newcommand{\kF}{k_{\mathrm{F}}}
\newcommand{\vF}{v_{\mathrm{F}}}
\newcommand{\wD}{\omega_{\mathrm{D}}}
\newcommand{\Tc}{T_{\mathrm{c}}}

\begin{document}

\title{
Nonlinear optical response of collective modes \\ in multiband superconductors assisted by nonmagnetic impurities
}

\date{\today}
\author{Yuta Murotani}
\affiliation{Department of Physics, The University of Tokyo, Tokyo 113-0033, Japan}
\author{Ryo Shimano}
\affiliation{Department of Physics, The University of Tokyo, Tokyo 113-0033, Japan}
\affiliation{Cryogenic Research Center, The University of Tokyo, Tokyo 113-0032, Japan}


\begin{abstract}
In multiband superconductors, multiple collective modes exist associated with the multiple order parameters.  
Oscillations of the amplitude and the relative phase of the order parameters are called Higgs and Leggett modes, respectively.
Recently, it has been suggested that nonmagnetic impurity scattering would enhance nonlinear coupling between the Higgs mode and 
an electromagnetic wave with a frequency located in the superconducting gap region, 
while its effect on the Leggett mode is still unresolved.
Here, we theoretically investigated the nonlinear optical response of multiband Bardeen-Cooper-Schrieffer-type superconductors in the presence of nonmagnetic impurities 
with a density matrix approach extending the Mattis-Bardeen model of linear response. 
We found that the drastic enhancement of nonlinear optical response due to the nonmagnetic impurity scattering occurs only for the Higgs modes and not for the Leggett mode.
As a result, both the light-induced dynamics of the superconducting gaps and the resulting third-harmonic generation are dominated by the Higgs modes.
We also examined the role of quasiparticle excitations to find that they give the subdominant contribution to the third-harmonic generation.
\end{abstract}

\maketitle

\section{Introduction}

Collective modes in superconductors have recently gained great interest owing to the development of nonlinear terahertz spectroscopy \cite{Matsunaga2017-1}. 
Because a superconducting order parameter is a complex quantity in general, 
one superconducting degree of freedom accommodates one amplitude mode \cite{Volkov1973} and one phase mode \cite{Nambu1960,Goldstone1961,Goldstone1962}. 
In superconductors, the Anderson-Higgs mechanism elevates the energy of the phase mode to the plasma frequency 
far larger than the superconducting gap \cite{Anderson1963,Higgs1964-1,Englert1964,Higgs1964-2,Guralnik1964}.
As a result, the amplitude mode is left stable in the low-energy region 
and is specially called a Higgs mode \cite{Higgs1964-2,Varma2002,Pekker2015}. 
The Higgs mode was initially identified in a superconductor with a coexisting charge-density wave order by Raman spectroscopy \cite{Sooryakumar1980,Littlewood1981,Measson2014}. 
Nonlinear terahertz spectroscopy has enabled the observation of the Higgs mode in conventional superconductors as well, 
in the form of a free or forced oscillation and the resulting third-harmonic generation (THG) \cite{Matsunaga2013,Matsunaga2014,Matsunaga2017-2}. 
This technique has been also applied to unconventional superconductors to investigate elementary excitations and paring mechanisms \cite{Katsumi2018,Chu2019}. 

Multiband superconductors can accommodate a variety of collective modes because of multiple degrees of freedom. 
For example, two interacting phase modes are transformed into a high-energy plasma oscillation and a low-energy gapped mode, called a Leggett mode \cite{Leggett1966,Sharapov2002,Burnell2010}. 
The Leggett mode has been observed in a two-gap superconductor MgB$_2$ by Raman spectroscopy \cite{Blumberg2007,Klein2010,Cea2016-2}. 
Multiple Higgs modes should also interact with each other through the interband paring interaction \cite{Murotani2017}, though it is not yet experimentally observed. 
Even more interestingly than the single-band cases, nonlinear terahertz spectroscopy offers a unique opportunity to investigate interaction among these collective modes,  
and the origin of paring in multiband superconductors \cite{Krull2016}. 
Recently, a terahertz pump-probe measurement on MgB$_2$ has been reported \cite{Giorgianni2019}, 
where the pump-induced oscillation of the transmitted probe electric field is attributed to the Leggett mode. 
However, the theoretical analysis presented there neglects the effect of impurity scattering, which may underestimate the contribution from the Higgs mode. 

In single-band superconductors, it has been theoretically shown that optical transitions mediated by nonmagnetic impurity scattering 
significantly enhances nonlinear optical response of the Higgs mode \cite{Jujo2015,Jujo2018,Silaev2019}. 
Typical multiband superconductors, such as MgB$_2$ and iron-based superconductors, actually often exhibit characteristics of impurity scattering 
in the energy region around the superconducting gaps even in the linear response \cite{Ortolani2008,Pimenov2013}. 
Therefore, it is indispensable to take into account the effects of impurities 
beyond the clean-limit analysis \cite{Tsuji2015,Cea2016-1,Murotani2017,Cea2018,Giorgianni2019} to consider nonlinear optical response of multiband superconductors. 
For that purpose, we extend the Mattis-Bardeen (MB) model for linear response of single-band Bardeen-Cooper-Schrieffer (BCS)-type superconductors \cite{Mattis1958,Zimmermann1991,Berlinsky1993} to a multiband system and nonlinear regime.
We found that optical response of the Leggett mode is insensitive to nonmagnetic impurity scattering. 
Consequently, magnitude of the light-induced oscillation of the Leggett mode is far smaller than that of the Higgs mode assisted by impurities. 
We also revealed that the Higgs mode contributes to THG dominating the contributions from the Leggett mode and quasiparticle excitations.  

This paper is organized as follows. 
Section II formulates the Hamiltonian which takes into account both impurity scattering and light-matter interaction, 
and introduces a density matrix to describe light-induced collective modes and nonlinear optical response.
Section III analyzes the linear response of a two-band superconductor. 
Section IV considers light-induced collective modes in the presence of impurity scattering. 
First, we confirm the effect of impurities in a single-band superconductor which was previously tested by Green's function method \cite{Jujo2015,Jujo2018,Silaev2019}.
We then examine the excitation of the Higgs and Leggett modes in a two-band superconductor for several excitation conditions. 
Section V discusses THG, decomposing contributions from the Higgs, Leggett modes and quasiparticles. 
Finally, Sec. VI summarizes the obtained results. 

\section{Density matrix formulation}

\subsection{Hamiltonian}

We extend the Mattis-Bardeen (MB) model \cite{Mattis1958} of light-matter interaction in superconductors to a multiband system. 
First, the noninteracting part of the Hamiltonian is written as 
\begin{align}
\mathcal{H}_0=\sum_{i\kk\sigma}\epsilon_{i\kk}c_{i\kk\sigma}^\dag c_{i\kk\sigma}, \label{H0}
\end{align}
where $c_{i\kk\sigma}^\dag$ ($c_{i\kk\sigma}$) creates (annihilates) an electron with the crystal momentum $\kk$ and the spin $\sigma$ 
in an energy band labeled by $i$. 
We assume the inversion symmetry of the system requiring $\epsilon_{i\kk}=\epsilon_{i(-\kk)}$.
In the following, we restrict ourselves to the simplest case where every band is well described by a parabolic dispersion relation, 
\begin{align}
\epsilon_{i\kk}=s_i\left(\frac{\kk^2}{2m_i}-\epsilon_{\mathrm{F}i}\right),~
s_i=
\begin{cases}
+1 & \mathrm{electron~band} \\
-1 & \mathrm{hole~band}
\end{cases}\label{ek}
\end{align}
with $m_i(>0)$ and $\epsilon_{\mathrm{F}i}(>0)$ being the effective mass and the Fermi energy, respectively. 
The band extremum, set at the $\Gamma$ point ($\kk=0$) in Eq. (\ref{ek}) for simplicity, can be moved to any point in the Brillouin zone, 
as long as the overall inversion symmetry is maintained. 

To be exact, $\kk$ cannot be regarded as the crystal momentum in the presence of impurities that break the translational symmetry of the crystal. 
Nevertheless, it serves as a quantum number labeling the true energy eigenstates. 
When the concentration of impurities is not too high, the energy eigenvalues should be hardly modified, which allows us to use Eq. (\ref{ek}) even for dirty systems. 

For later convenience, the Fermi momentum $k_{\mathrm{F}i}=\sqrt{2m_i\epsilon_{\mathrm{F}i}}$ 
and the Fermi velocity $v_{\mathrm{F}i}=k_{\mathrm{F}i}/m_i$ are introduced.
Note that we have chosen $\hbar=1$.
We assume band-dependent quantities (e.g., $\epsilon_{\mathrm{F}i}$ and $k_{\mathrm{F}i}$) to be in the same order for every band. 
The respective order will be represented by symbols without the band index (e.g, $\eF$ and $\kF$).

Next, the pairing interaction is given by 
\begin{align}
\mathcal{H}_{\mathrm{int}}&=-\sum_{ij}U_{ij}\sum_{\kk\kk'}c_{i\kk\uparrow}^\dag c_{i(-\kk)\downarrow}^\dag c_{j(-\kk')\downarrow}c_{j\kk'\uparrow}, \label{Hint}
\end{align}
where $U_{ij}$ is the pairing potential between $i$th and $j$th bands \cite{Suhl1959}. 
Throughout the paper, the volume of the system is set to unity.

In single-band superconductors, Anderson's theorem guarantees robustness of the paring potential (\ref{Hint}) against nonmagnetic impurities \cite{Anderson1959}. 
In multiband systems, however, Anderson's theorem generally fails because of interband scattering \cite{Stanev2014}. 
As a result, $U_{ij}$ can depend on density of impurities and concomitantly on momenta $\kk$ and $\kk'$. 
It then should modify the transition temperature and the magnitude of superconducting gaps \cite{Golubov1997}. 
However, it has been shown that both quantities only slightly depend on the concentration of impurities in MgB$_2$ \cite{Mazin2002,Ortolani2008}. 
In FeSe being another typical multiband superconductor, 
superconductivity is more robust against impurities than expected from theory considering interband impurity scattering \cite{Urata2016}. 
Therefore, we neglect this effect and adopt Eq. (\ref{Hint}) in the following. 
Physically, this assumption requires that the impurities cannot scatter electrons over the separation between different Fermi surfaces. 

According to the above reasoning, we neglect interband impurity scattering also in optical transitions. 
Then, only intraband transitions suffice for consideration of low-energy optical response, 
described by an interaction Hamiltonian 
\begin{align}
\mathcal{H}_1=-\sum_{i\kk\kk'\sigma}\left(\J_{i\kk\kk'}\cdot\textbf{A}\right)c_{i\kk\sigma}^\dag c_{i\kk'\sigma}.\label{H1} 
\end{align}
Here, $\J_{i\kk\kk'}$ is the matrix element of the current operator $e\hat{\mathbf{p}}/m_0$ 
with the momentum operator $\hat{\mathbf{p}}$, the charge $e(<0)$ and the mass $m_0$ of a free electron, 
and $\textbf{A}$ is vector potential of the external light field assumed to be spatially homogeneous upon dipole approximation. 
We also have to take into account the nonlinear coupling to the light field, 
\begin{align}
\mathcal{H}_2&=\sum_{i\kk\sigma}s_i\left(\frac{e^2\textbf{A}^2}{2m_i}\right)c_{i\kk\sigma}^\dag c_{i\kk\sigma}, \label{H2}
\end{align}
to preserve the gauge invariance of the model. 
Finally, the total Hamiltonian is given by 
\begin{align}
\mathcal{H}=\mathcal{H}_0+\mathcal{H}_1+\mathcal{H}_2+\mathcal{H}_{\mathrm{int}}. \label{Ham}
\end{align}

$\mathcal{H}_1$ and $\mathcal{H}_2$ describe the paramagnetic and diamagnetic coupling to the light field, respectively. 
In an ideal crystal (the clean limit explained below), 
the former coupling is known to be negligible \cite{Matsunaga2014,Tsuji2015}. 
In dirty systems, however, the paramagnetic coupling is essential. 
For example, $\mathcal{H}_1$ is indispensable to describe linear response of dirty single-band BCS superconductors \cite{Mattis1958,Zimmermann1991,Berlinsky1993}. 
It has been predicted that this coupling would dominate also the nonlinear optical response \cite{Jujo2015,Jujo2018,Yu2017,Yang2018a,Yang2018b,Silaev2019}. 
In terms of the Green's function method, this is because impurity scattering produces nonvanishing Feynman diagrams that vanish in the clean limit \cite{Tsuji2016}. 
We adopt the MB model rather than employing Green's functions to examine effects of the paramagnetic coupling assisted by the impurity scattering.

The essence of the MB model \cite{Mattis1958} lies in assuming the transition matrix element to be well approximated by a Lorentzian function, 
\begin{align}
\left\langle\left|\mathbf{e}\cdot\J_{i\kk\kk'}\right|^2\right\rangle_{\mathrm{Av}}&\equiv\iint\frac{d\Omega_{\kk}}{4\pi}\frac{d\Omega_{\kk'}}{4\pi}\left|\mathbf{e}\cdot\J_{i\kk\kk'}\right|^2\nonumber\\
&\simeq\frac{(ev_{\mathrm{F}i})^2}{3N_i(0)}W_i(\epsilon_{i\kk},\epsilon_{i\kk'}),\label{MB1}\\
W_i(\epsilon,\epsilon')&=\frac{1}{\pi}\frac{\gamma_i}{(\epsilon-\epsilon')^2+\gamma_i^2},\label{MB2}
\end{align}
where $\mathbf{e}$ is the polarization vector of light ($\A(t)=A(t)\mathbf{e}$), 
$N_i(0)=m_ik_{\mathrm{F}i}/2\pi^2$ is the density of states per spin on the Fermi surface,
and $\gamma_i$ is the impurity scattering rate. 
See Appendix \ref{matrix_element} for derivation. Here, it is assumed that
\begin{align}
\gamma\ll\eF.\label{condition1}
\end{align}
Condition (\ref{condition1}) requires the impurity scattering to occur only within a thin shell around the Fermi surfaces. 

When the considered energy region is far smaller than $\gamma$ (the dirty limit), the transition matrix element can be approximated by an energy-independent constant.
Such an approach has been used to calculate transient optical conductivity of photoexcited superconductors \cite{Chou2017,Kennes2017}.

\subsection{Energy scales}

The interaction Hamiltonian $\mathcal{H}_{\mathrm{int}}$ is treated in the mean-field approximation. 
As a result, the superconducting gap function
\begin{align}
\Delta_i&=\sum_jU_{ij}\sum_{\kk}\langle c_{j(-\kk)\downarrow}c_{j\kk\uparrow}\rangle,\label{self_consistency}
\end{align}
is defined and regarded as the multicomponent order parameter. 
At the thermal equilibrium, $2\Delta_i^{\mathrm{eq}}$ is self-consistently determined and serves as a gap in the excitation spectrum (see Appendix \ref{quasiparticle_representation}).
As in usual BCS theory, we assume 
\begin{align}
2\Delta^{\mathrm{eq}}\ll\eF.\label{condition2}
\end{align}

Now, we have three characteristic energy scales: \\
(A) The Fermi energy $\eF$. \\
(B) The scattering rate $\gamma$. \\
(C) The superconducting gap 2$\Delta^{\mathrm{eq}}$. \\
We have already assumed two inequalities, namely Eq. (\ref{condition1}) between (A) and (B) and Eq. (\ref{condition2}) between (A) and (C). 
The remaining relationship between (B) and (C) concerns the distinction between ``clean'' and ``dirty'' systems:
\begin{align}
\begin{cases}
\gamma\ll2\Delta^{\mathrm{eq}} & \mathrm{clean~limit}, \\
\gamma\gg2\Delta^{\mathrm{eq}} & \mathrm{dirty~limit}.
\end{cases}
\end{align}
In the clean limit, we can neglect the impurity scattering in the frequency region around $2\Delta^{\mathrm{eq}}$ that we are interested in. 
On the other hand, the impurity scattering plays a significant role in mediating nonlinear interaction between light and collective modes in the dirty limit \cite{Jujo2015,Jujo2018}.
We also call more general cases ``dirty,'' when $\gamma$ and $2\Delta^{\mathrm{eq}}$ are in the same order.

\subsection{Equation of motion}

To consider linear and nonlinear optical response, we introduce the density matrix for Bogoliubov quasiparticles, 
\begin{align}
\rho_{i\kk\kk'}^{ab}=\langle\psi_{i\kk}^{a\dag}\psi_{i\kk'}^b\rangle\quad(a,b=1,2), \label{density_matrix}
\end{align}
where the two-component spinor $\psi_{i\kk}$ is defined by Eq. (\ref{spinor}) in Appendix \ref{quasiparticle_representation}.
We solve the corresponding equation of motion, i.e., Eq. (\ref{EoM}) in Appendix \ref{quasiparticle_representation}, in a perturbative manner with respect to the external field $\A(t)$: 
\begin{align}
\rho_{i\kk\kk'}^{ab}&=\rho_{i\kk\kk'}^{ab}\Big|_0+\rho_{i\kk\kk'}^{ab}\Big|_1+\rho_{i\kk\kk'}^{ab}\Big|_2+\cdots, 
\end{align}
where the additional subscript denotes the order of $\A$. 

Motion of the gap function can be calculated through Eq. (\ref{self_consistency}).
Here, one has to pay attention to consistency of the formulation.
The variation
$\delta\Delta_i(t)=\Delta_i(t)-\Delta_i^{\mathrm{eq}}$ formally acts as an external field; the explicit perturbation Hamiltonian is given by Eq. (\ref{dHint}) in Appendix \ref{quasiparticle_representation}.
The resulting motion of quasiparticles in turn affects $\delta\Delta_i(t)$ itself through the definition (\ref{self_consistency}).
This feedback process induces the collective modes and must be taken into account in numerical calculations.

To consider optical response itself, we need to calculate electric current density
\begin{align}
\textbf{j}&=-\left\langle\frac{\delta\mathcal{H}}{\delta\A}\right\rangle=\textbf{j}_{\mathrm{P}}+\textbf{j}_{\mathrm{D}},
\end{align}
with the paramagnetic component $\textbf{j}_{\mathrm{P}}=-\langle\delta\mathcal{H}_1/\delta\A\rangle$
and the diamagnetic component $\textbf{j}_{\mathrm{D}}=-\langle\delta\mathcal{H}_2/\delta\A\rangle$.
Their explicit expressions are given by Eqs. (\ref{jP}), (\ref{jD}) in Appendix \ref{quasiparticle_representation}.

\section{Linear response}\label{SecIII}

First, we consider the linear response. 
It is known that the collective modes do not respond to light in the linear response regime in the absence of any other external fields. 
As a result, we can neglect $\delta\mathcal{H}_{\mathrm{int}}$ and consider only $\mathcal{H}_1$ as the perturbation Hamiltonian.
We find that the paramagnetic component is given by
\begin{align}
&\jj_{\mathrm{P}}(t)\Big|_1=\mathbf{e}\sum_{i}\frac{e^2n_i}{m_i}\iint d\epsilon~d\epsilon'~W_i(\epsilon,\epsilon')\nonumber\\
&\times\left[l_{i}(\epsilon,\epsilon')^2\operatorname{Re}F_i^{11}(\epsilon,\epsilon')+p_{i}(\epsilon,\epsilon')^2\operatorname{Re}F_i^{21}(\epsilon,\epsilon')\right], \label{jP1-2}
\end{align}
where $n_i=k_{\mathrm{F}i}^3/3\pi^2$ is the density of carriers (either electrons or holes), 
$l_i(\epsilon_{i\kk},\epsilon_{i\kk'})=l_{i\kk\kk'}$ and $p_i(\epsilon_{i\kk},\epsilon_{i\kk'})=p_{i\kk\kk'}$ are coherence factors defined by Eq. (\ref{coherence_factors}),
and the function $F_{i}^{ab}(\epsilon,\epsilon')$ follows
\begin{align}
\left[i\frac{\partial}{\partial t}-(E'-E)\right]F_{i}^{11}(\epsilon,\epsilon')&=(f'-f)A, \label{EoM1-1}\\
\left[i\frac{\partial}{\partial t}-(E'+E)\right]F_{i}^{21}(\epsilon,\epsilon')&=-(1-f-f')A, \label{EoM1-2}\\
F_{i}^{22}(\epsilon,\epsilon')=F_{i}^{11}(\epsilon,\epsilon')^*,~F_{i}^{12}&(\epsilon,\epsilon')=F_{i}^{21}(\epsilon,\epsilon')^*,\label{EoM1-3}
\end{align}
with $E=E_{i\kk}$, $f'=f_{i\kk'}$, etc. 
On the other hand, the diamagnetic component is given by 
\begin{align}
\jj_{\mathrm{D}}(t)\Big|_1&=-\A\sum_{i}\frac{e^2n_i}{m_i}. \label{jD1-1}
\end{align}
For derivation, see Appendix \ref{linear_response}.

After the summation $\jj(t)|_1=\jj_{\mathrm{P}}(t)|_1+\jj_{\mathrm{D}}(t)|_1$, we can obtain the optical conductivity 
\begin{align}
\sigma(\omega)&=\sigma_1(\omega)+i\sigma_2(\omega)=\frac{j(\omega)}{E(\omega)}, \label{OC}
\end{align}
where $j(\omega)$ is the Fourier transform of $\textbf{e}\cdot\jj(t)|_1$ and
$E(\omega)$ is that of the electric field $E(t)=-\partial A(t)/\partial t$. 
The real part $\sigma_1(\omega)$ corresponds to absorption of light 
while the imaginary part $\sigma_2(\omega)$ corresponds to refraction.

For simplicity, we considered a two-band system with the following parameters: \\
(Band 1) $s_1=+1$, $m_1=1.0$, $\epsilon_{\mathrm{F}1}=500$, $\gamma_1=10$. \\
(Band 2) $s_2=-1$, $m_2=1.2$, $\epsilon_{\mathrm{F}2}=300$, $\gamma_2=10$. \\
(Paring interaction) $U_{11}=0.08$, $U_{22}=0.18$, $U_{12}=0.05$, $\wD=10$.

The chosen parameters qualitatively simulate MgB$_2$ with electron- and hole-like bands 
with $\eF\sim1$ eV, $\gamma\sim0.1$ eV, and $2\Delta\sim0.01$ eV \cite{Ortolani2008}.
To avoid formal failure of the weak-coupling approximation in our theory, we chose relatively small values for the paring potential 
such that $N_2(0)U_{22}\simeq0.3$ while the reported values are relatively large, e.g., $N_2(0)U_{22}\simeq1$ \cite{Blumberg2007}.
We expect that the results presented below will be qualitatively the same even for strong couplings.

\begin{figure}
\centering
\includegraphics[width=8.5cm]{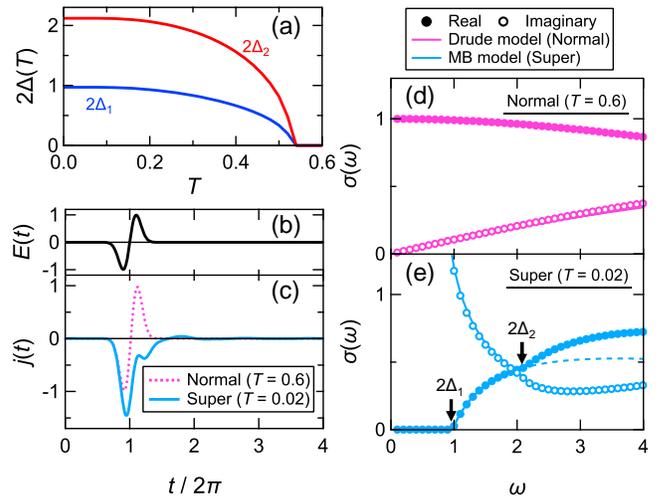}
\caption{Linear response of  a two-band BCS superconductor. 
(a) Temperature dependence of the superconducting gaps considered in this work. 
(b) Electric field and (c) the induced electric current above $\Tc$ (dotted) and below $\Tc$ (solid). 
(d) Optical conductivity above $\Tc$. The solid and open circles correspond to the simulated real and imaginary parts, respectively. Solid lines are Drude model. 
(e) Optical conductivity below $\Tc$. Solid lines are MB model. Dashed line shows the contribution to the real part from the band with the lower gap.
}\label{Fig1}
\end{figure}

Temperature dependence of the superconducting gaps obtained from the gap equation (\ref{gap_eqn}) is shown in Fig. \ref{Fig1} (a). 
Both bands establish superconductivity ($2\Delta_i\neq0$ with the superscript ``eq'' omitted) at a common critical temperature $\Tc=0.54$. 
They take the maximum values at the absolute zero, $2\Delta_1(0)=0.97$ and $2\Delta_2(0)=2.1$.

We introduced a monocycle electric field described by 
\begin{align}
A(t)=A_0\exp\left(-\frac{t^2}{2\tau^2}\right)\cos\Omega t,\label{vector_potential}
\end{align}
with $\tau=2\pi\times0.1$, $\Omega=0$. 
The electric field waveform is plotted in Fig. \ref{Fig1} (b). 
We solved Eqs. (\ref{EoM1-1}) and (\ref{EoM1-2}) in the time domain to calculate the electric current. 
The simulated results are plotted in Fig. \ref{Fig1} (c) for $T=0.6>\Tc$ (dotted) and $T=0.02<\Tc$ (solid). 
While the former closely follows the electric field waveform, the latter shows a characteristic oscillatory structure. 
We calculated the optical conductivity (\ref{OC}) and show it in Fig. 1 (d) for $T=0.6$ and in (e) for $T=0.02$. 
The solid and open circles are the real and imaginary parts, respectively. 
Above $\Tc$, the optical conductivity is structureless in the plotted frequency range as seen in (d). 
It can be shown that the response exactly coincides with the Drude model \cite{Mattis1958}, 
\begin{align}
\sigma_{\mathrm{Drude}}(\omega)=\sum_i\frac{ie^2n_i}{m_i(\omega+i\gamma_i)}, 
\end{align}
which is shown in Fig. (d) as solid lines. 
Small discrepancy between the calculated result and the Drude model in $\sigma_2(\omega)$ arises from the cutoff introduced in numerical integration. 

On the other hand, below $\Tc$ (Fig. \ref{Fig1} (e)), 
the imaginary part (open circles) diverges toward $\omega\to0$ being an indicator of superconductivity. 
Even more remarkably, the real part (solid circles) shows a double-gapped structure. 
The absorption edges coincide with the superconducting gaps $2\Delta_i$ indicated by arrows. 
The dashed line depicts the contribution from the band with the smaller gap $2\Delta_1$ as a guide to the eye. 
It can be shown that the response exactly coincides with the sum of MB conductivity explicitly given in Refs. \cite{Zimmermann1991,Berlinsky1993} over all bands. 
MB model conductivity is plotted as the solid lines in Fig. \ref{Fig1} (e), displaying a good agreement with the simulation. 
This ensures the validity of our time-domain calculation.

\section{Nonlinear excitation of collective modes}\label{SecIV}

Let us proceed to the next order. 
The relevant equation of motion is given in Appendix \ref{2nd-order}.
It is well known that an isotropic system exhibits no even-order nonlinearities, so that 
\begin{align}
\jj(t)\Big|_2=0.
\end{align}
Therefore, we can concentrate on $\delta\Delta_i(t)|_2$.
Assuming the particle-hole symmetry, the real and imaginary parts are rewritten as 
\begin{align}
\delta\Delta_i'\Big|_2&=\sum_jU_{ij}N_j(0)\int d\epsilon\left\{-u_{j}(\epsilon)v_{j}(\epsilon)\left[r_{j}^{11}(\epsilon)-r_{j}^{22}(\epsilon)\right]\right.\nonumber\\
&\quad\left.+\frac{1}{2}[u_{j}(\epsilon)^2-v_{j}(\epsilon)^2][r_{j}^{21}(\epsilon)+r_{j}^{12}(\epsilon)]\right\},\label{real_part2}\\
\delta\Delta_i''\Big|_2&=\sum_jU_{ij}N_j(0)\int d\epsilon~\frac{1}{2i}[r_{j}^{21}(\epsilon)-r_{j}^{12}(\epsilon)],\label{imaginary_part2}
\end{align}
respectively, 
where $u_i(\epsilon_{i\kk})=u_{i\kk}$, $v_i(\epsilon_{i\kk})=v_{i\kk}$, and
\begin{align}
r_i^{ab}(\epsilon_{i\kk})=\int\frac{d\Omega_{\kk}}{4\pi}\rho_{i\kk\kk}^{ab}\Big|_2.
\end{align}
Thus we can drop the dependence of the density matrix on the angle of $\kk$. 
The induced motion of the density matrix can be also decomposed into the quasiparticle, Higgs mode, and Leggett mode.

\subsubsection{Quasiparticles}
The diagonal components $r_i^{11}(\epsilon)$ and $r_i^{22}(\epsilon)$ correspond to the quasiparticle excitation. 
They follow 
\begin{align}
i\frac{\partial}{\partial t}r_i^{11}(\epsilon)&=S_i^{11}(\epsilon),\label{EoM2-1}\\
r_i^{22}(\epsilon)&=-r_i^{11}(\epsilon),
\end{align}
where 
\begin{align}
S_i^{11}(\epsilon)&=-2iA\frac{(ev_{\mathrm{F}i})^2}{3}\int d\epsilon'\left[l_{i}(\epsilon,\epsilon')^2\operatorname{Im}F_i^{11}(\epsilon,\epsilon')\right.\nonumber\\
&\quad\left.-p_i(\epsilon,\epsilon')^2\operatorname{Im}F_i^{21}(\epsilon,\epsilon')\right]W_i(\epsilon,\epsilon').\label{S11}
\end{align}
Because $S_i^{11}(\epsilon)=S_i^{11}(-\epsilon)$, we have 
\begin{align}
r_i^{11}(\epsilon)=r_i^{11}(-\epsilon).
\end{align}

\subsubsection{Higgs mode}

The non-diagonal components $r_i^{21}(\epsilon)=r_i^{12}(\epsilon)^*$ correspond to the collective modes. 
To separate the Higgs mode and the Leggett mode, we further decompose them into the odd and even parts, 
\begin{align}
r_i^{21}(\epsilon)&=r_i^{21,\mathrm{odd}}(\epsilon)+r_i^{21,\mathrm{even}}(\epsilon).
\end{align}
Each of them satisfies 
\begin{align}
r_i^{21,\mathrm{odd}}(\epsilon)&=-r_i^{21,\mathrm{odd}}(-\epsilon),\\
r_i^{21,\mathrm{even}}(\epsilon)&=r_i^{21,\mathrm{even}}(-\epsilon).
\end{align}
Among them, the odd component corresponds to the Higgs mode. 
This can be easily understood because only the odd component contributes to the right-hand side of Eq. (\ref{real_part2}). 
Note that $r_i^{11}(\epsilon)$ and $r_i^{22}(\epsilon)$ also contribute, but because their motion is determined by a closed equation of motion (\ref{EoM2-1}), 
quasiparticle excitations only trigger the Higgs mode without any feedback effect. 
The odd component $r_i^{21,\mathrm{odd}}(\epsilon)$ follows 
\begin{align}
\left(i\frac{\partial}{\partial t}-2E\right)r_i^{21,\mathrm{odd}}(\epsilon)&=-(1-2f)(u^2-v^2)\delta\Delta_i'\Big|_2\nonumber\\
&\quad+S_i^{21}(\epsilon), \label{EoM2-2}
\end{align}
where 
\begin{align}
S_i^{21}(\epsilon)&=-2A\frac{(ev_{\mathrm{F}i})^2}{3}\int d\epsilon'~W_i(\epsilon,\epsilon')l_{i}(\epsilon,\epsilon')p_{i}(\epsilon,\epsilon')\nonumber\\
&\quad\times\left[F_i^{21}(\epsilon,\epsilon')-F_i^{22}(\epsilon,\epsilon')\right] \label{S21}\\
&=-S_i^{21}(-\epsilon).\nonumber
\end{align}
The source term $S_i^{21}(\epsilon)$ arises from the paramagnetic coupling with light mediated by the impurity scattering. 
Note that Eq. (\ref{EoM2-2}) has to be solved keeping Eq. (\ref{real_part2}) to be always satisfied. 
This self-consistency condition induces the Higgs mode resonance.

\begin{table}
\centering
\begin{tabular}{@{\hspace{3mm}}c@{\hspace{3mm}}||@{\hspace{3mm}}c@{\hspace{3mm}}|@{\hspace{3mm}}c@{\hspace{3mm}}} \hline
Channel & Para ($\pp\cdot\A$) & Dia ($\A^2$) \\ \hline
HM ($\delta\Delta'$) & $\eF\gamma/\Delta^2~\to~\eF/\gamma$ & 0 [$\Delta/\eF$] \\ 
LM ($\delta\Delta''$) & $\gamma/\Delta~\to~\Delta/\gamma~(*)$ & 1 \\ \hline
\end{tabular}
\caption{Order of the light-induced collective modes, in the unit of the ponderomotive energy $e^2\A^2/2m$. 
Arrows connects the clean limit ($\gamma\ll2\Delta$) on the left side and the dirty limit ($\gamma\gg2\Delta$) on the right side. 
For a reference, square brackets show the case when the energy dispersion relation exhibits a nonparabolicity, calculated in the clean limit \cite{Tsuji2015}. 
The asterisked part is neglected in the main text.
}\label{TableI}
\end{table}

For the incident frequency in the order of $2\Delta$, the amplitude of the induced Higgs mode is estimated as follows (see Appendix \ref{order_estimation}):  
\begin{align}
\delta\Delta'\sim\frac{e^2\A^2}{2m}\times
\begin{cases}
(\eF\gamma/\Delta^2) & \gamma\ll2\Delta, \\
(\eF/\gamma) & \gamma\gg2\Delta.
\end{cases}\quad\mathrm{(Para)}\label{HM_Para}
\end{align}
This estimation predicts the most efficient excitation of the Higgs mode at $\gamma\sim2\Delta$, which will be confirmed numerically. 
Equation (\ref{HM_Para}) arises from only the paramagnetic coupling (abbreviated as ``Para'') because 
the diamagnetic coupling (to be abbreviated as ``Dia'') vanishes in Eq. (\ref{EoM2-2}). 
However, it is known that the latter can also induce the Higgs mode when the energy dispersion relation exhibits a nonparabolicity in the clean limit \cite{Tsuji2015}. 
Because the diamagnetic term (\ref{H1}) of the Hamiltonian is less sensitive to impurity scattering than the paramagnetic term (\ref{H2}), 
it is reasonable to assume that the correction by a small nonparabolicity does not depend on $\gamma$: 
\begin{align}
\delta\Delta'\sim\frac{e^2\A^2}{2m}\times\frac{\Delta}{\eF}.\quad\mathrm{(Dia)} \label{HM_Dia}
\end{align}
Here, we have used that $U\sim\eF$ which is valid for phonon-mediated interactions \cite{Coleman2015}.
Because the diamagnetic contribution (\ref{HM_Dia}) is smaller than paramagnetic one (\ref{HM_Para}), 
the paramagnetic coupling is more important in optical excitation of the Higgs mode. 
Result of the order estimation is summarized in the second row of Table \ref{TableI}.

The excitation mechanism of the Higgs mode can be understood as follows. 
Because this mode has an even parity in momentum space, 
dipole-allowed intermediate states are necessary to excite it through a two-photon process via the paramagnetic coupling. 
In the clean limit, there is no dipole-allowed excitation because all the spectral weight concentrates on the zero frequency.
However, presence of impurity scattering produces dipole-allowed excitations as exemplified by Fig. \ref{Fig1} (e), 
which then enhance the two-photon excitation of the Higgs mode \cite{Jujo2015}.

It has been suggested that retardation of the phonon-mediated interaction also enhances the nonlinear optical response of the Higgs mode \cite{Tsuji2016}. 
Even in this case, the enhancement arises from the paramagnetic coupling, termed there ``resonant coupling'' from an analogy with Raman scattering.
It is an interesting problem to compare it with the effect of impurity scattering quantitatively, which is outside the scope of this paper. 
However, retarded interaction predicts comparable contributions by the Higgs mode induced via the paramagnetic or ``resonant'' coupling and by quasiparticles excited via the diamagnetic or ``nonresonant'' coupling in THG. 
This contrasts with our discussion in the next section which predicts dominance of the former contribution. 
Therefore, it is possible that impurity scattering is more important in enhancing the coupling between the Higgs mode and light. 

\subsubsection{Leggett mode}

The even component $r_i^{21,\mathrm{even}}(\epsilon)$ corresponds to the Leggett mode. 
It follows 
\begin{align}
\left(i\frac{\partial}{\partial t}-2E\right)r_i^{21,\mathrm{even}}(\epsilon)&=(1-2f)2uv\left(s_i\frac{e^2\A^2}{2m_i}\right)\nonumber\\
&\quad-i(1-2f)\delta\Delta_i''\Big|_2.\label{EoM2-3}
\end{align}
The paramagnetic coupling vanishes here. 
This equation has to be solved consistently with Eq. (\ref{imaginary_part2}), which induces the Leggett mode. 
Its amplitude is estimated as 
\begin{align}
\delta\Delta''\sim\frac{e^2\A^2}{2m}.\quad\mathrm{(Dia)}\label{LM_Dia}
\end{align}
This result is corrected by a particle-hole asymmetry which modifies energy integrals 
through the finite slope of the density of states on the Fermi surface.
The resulting correction is in the following order: 
\begin{align}
\delta\Delta''\sim\frac{e^2\A^2}{2m}\times
\begin{cases}
(\gamma/\Delta) & \gamma\ll2\Delta, \\
(\Delta/\gamma) & \gamma\gg2\Delta.
\end{cases}\quad\mathrm{(Para)}\label{LM_Para}
\end{align}
This correction is not essential, because it does not exceed Eq. (\ref{LM_Dia}). 
Therefore, we will neglect it in the following. 
Result of the order estimation is summarized in the third row of Table \ref{TableI}. 

The diamagnetic contribution (\ref{LM_Dia}) 
has been explained in terms of the coupling between the Leggett mode and the potential difference between Fermi surfaces induced by the diamagnetic coupling \cite{Murotani2017}. 
Along this line, one can infer from the above result that quasiparticle excitation caused by the paramagnetic coupling 
induces no potential difference in the presence of particle-hole symmetry. 
This resembles the behavior of chemical potential being independent of temperature when carriers exhibit a constant density of states.
For a system slightly lacking the particle-hole symmetry, the paramagnetic contribution (\ref{LM_Para}) 
can be viewed as a two-photon excitation similar to that of the Higgs mode.


\subsection{Single-band case}

\begin{figure}
\centering
\includegraphics[width=8.5cm]{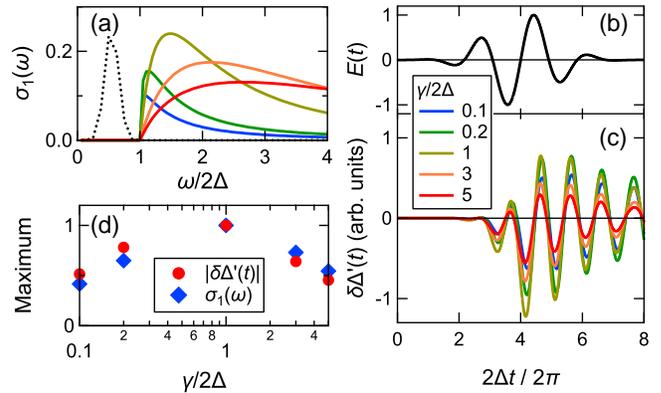}
\caption{Optical response of a single-band BCS superconductor at $T=0$, dependent on the scattering rate $\gamma$. 
(a) The real part of the optical conductivity $\sigma_1(\omega)$ for $\gamma/2\Delta=0.1$ (blue), 0.2 (green), 1 (yellow), 3 (orange) and 5 (red). 
Pump power spectrum is also shown as the dotted line, whose electric field waveform is plotted in (b).
(c) Forced oscillation of the superconducting gap $\delta\Delta'(t)$. Colors are the same as in (a). 
(d) $\gamma$-dependence of the maximum of $|\delta\Delta'(t)|$ (circles) and $\sigma_1(\omega)$ (squares), normalized to 1 at $\gamma/2\Delta=1$.
}\label{Fig2}
\end{figure}

Before discussing the multiband case, let us verify the effect of impurity scattering on the Higgs mode in a single-band superconductor, 
previously investigated by the Green's function method \cite{Jujo2015}. 
We used $N(0)U=0.27$ and $\wD=10$, which leads to $2\Delta=1$ at $T=0$. 
The real part of the optical conductivity $\sigma_1(\omega)$ is plotted in Fig. \ref{Fig2} (a) 
for $\gamma/2\Delta$=0.1, 0.2, 1, 3, and 5, as blue, green, yellow, orange, and red lines, respectively. 
We introduced a multicycle pulse described by vector potential (\ref{vector_potential}) with $\tau=2\pi\times1$, $\Omega=0.5$.
Its waveform $E(t)$ is shown in Fig. \ref{Fig2} (b) 
while the power spectrum $|E(\omega)|^2$ is plotted as the dotted line in (a).
The latter is located inside the gap $2\Delta$, 
indicating no optical excitation of quasiparticles. 
The simulated dynamics of $\delta\Delta'(t)$ is plotted in Fig. \ref{Fig2} (c) with the same parameters and colors as in (a). 
All curves show a clear oscillation with the doubled frequency $2\Omega$. 
Free oscillation of the Higgs mode remains after excitation ($2\Delta t/2\pi>6$), 
because the incident pulse satisfies the resonance condition for the Higgs mode, $2\Omega=2\Delta$ \cite{Murotani2017,Jujo2018}. 
The amplitude of oscillation, however, depends on $\gamma$. 
We plotted the maximum of $|\delta\Delta'(t)|$ as a function of $\gamma$ as circles in Fig. \ref{Fig2} (d), 
which takes the maximum value at $\gamma\sim2\Delta$ consistently with the above order estimation. 
In Fig. \ref{Fig2} (d), we also plotted the maximum value of $\sigma_1(\omega)$, 
that clearly correlates with the amplitude of the Higgs mode. 
This result indicates that virtual excitation of optically active intermediate states concerns excitation of the Higgs mode. 
All these properties successfully reproduce the results obtained by the Green's function method \cite{Jujo2015}.

\subsection{Two-band case}

\begin{figure}
\centering
\includegraphics[width=8.5cm]{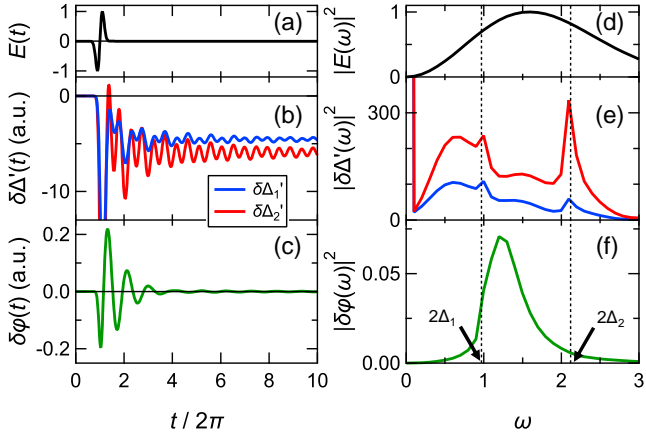}
\caption{Nonadiabatic excitation of a two-band BCS superconductor.
(a) Electric field waveform. 
(b) Induced Higgs mode. The upper and lower lines correspond to the smaller and larger gaps, respectively. a.u. stands for arbitrary units.
(c) Induced Leggett mode, shown in a form of the phase difference $\delta\varphi(t)=\delta\Delta_1''(t)/\Delta_1-\delta\Delta_2''(t)/\Delta_2$. 
(d)-(f) Power spectrum of the electric field, the Higgs mode, and the Leggett mode, respectively. 
The position of superconducting gaps $2\Delta_{1,2}$ are shown by dotted lines.
}\label{Fig3}
\end{figure}

Now, we return to the two-band system considered in the last section. 
We concentrate on $T=0.02$ well below $\Tc$. 
First, we examine a nonadiabatic excitation in which the electric field varies faster than the superconducting response time $2\pi/2\Delta$. 
The monocycle pulse used in the previous section fits this purpose. 
In Fig. \ref{Fig3} (a), the electric field waveform is shown again. 
Its power spectrum plotted in Fig. \ref{Fig3} (d) exhibits a broad bandwidth covering both gaps indicated by dotted lines. 
Figure \ref{Fig3} (b) shows the induced dynamics of $\delta\Delta_{1,2}'(t)$. 
Oscillations after excitation correspond to the Higgs mode. 
Both variables asymptotically approach negative values because excitation of Bogoliubov quasiparticles by a broadband pulse suppresses the superconducting order. 
Figure \ref{Fig3} (e) shows the power spectra of $\delta\Delta_{1,2}'(t)$. 
Both of them exhibit peaks at $\omega=2\Delta_1$ and $2\Delta_2$, being the resonance energies of the two Higgs modes \cite{Murotani2017}. 
Due to the interband interaction $U_{12}$, they mutually interact  so that both gaps oscillate with both frequencies. 
On the other hand, Fig. \ref{Fig3} (c) shows dynamics of the phase difference
\begin{align}
\delta\varphi(t)&=\frac{\delta\Delta_1''(t)}{\Delta_1}-\frac{\delta\Delta_2''(t)}{\Delta_2}, 
\end{align}
which displays a damped oscillation of the Leggett mode. 
Its power spectrum is shown in Fig. \ref{Fig3} (f). 
Because relatively large interband interaction $U_{12}$ pushes up the resonance energy above the lower gap $2\Delta_1$, 
the mode acquires finite lifetime coming from decay into the quasiparticle continuum. 
These oscillations of the gaps may be detectable in pump-probe experiments \cite{Matsunaga2013}.
However, having seen that the Leggett mode is far smaller than the Higgs mode, $\delta\Delta''(t)\ll\delta\Delta'(t)$, the 
observation of the Leggett mode is expected to be difficult.

\begin{figure}
\centering
\includegraphics[width=8.5cm]{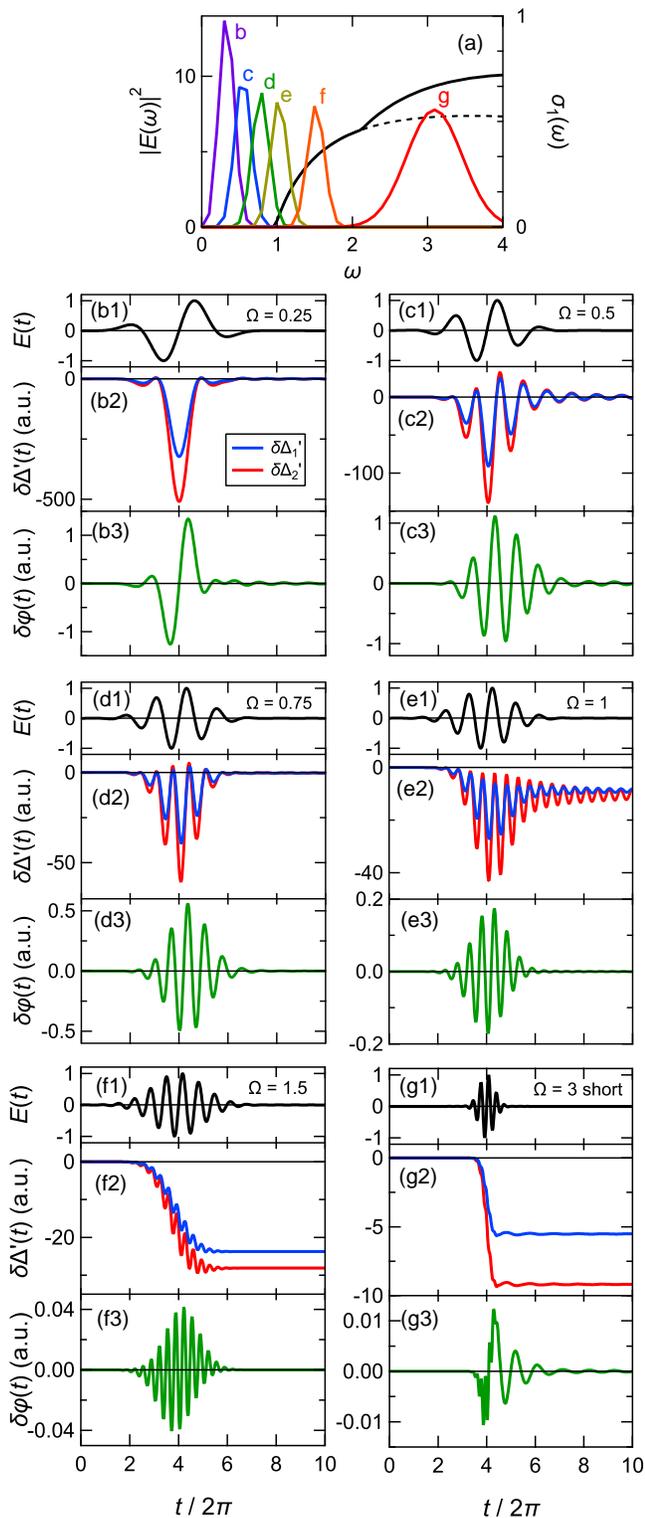}
\caption{Multicycle excitation of a two-band BCS superconductor.
(a) Power spectra of electric field with $\Omega=0.25$, 0.5, 0.75, 1, 1.5, and 3 from left to right, 
each corresponding to (b)-(g) respectively.
Pulse width is $\tau=2\pi\times1$ for (b)-(f) and $\tau=2\pi\times0.3$ for (g).
$\sigma_1(\omega)$ is also drawn. 
(b1)-(g1) Electric field waveform for $\Omega=0.25$, 0.5, 0.75, 1, 1.5, and 3, respectively. 
(b2)-(g2) Dynamics of $\delta\Delta_{1,2}'(t)$ for each excitation pulse. a.u. stands for arbitrary units.
(b3)-(g3) Dynamics of the phase difference $\delta\varphi(t)$.
}\label{Fig4}
\end{figure}

Next, we consider the excitation by multicycle pulses. 
In Fig. \ref{Fig4} (a), we plot the power spectra of pulses with $\Omega=0.25$, 0.5, 0.75, 1, 1.5, and 3, from left to right. 
Another parameter $\tau$ is set to $\tau=2\pi\times1$ for the former five and $\tau=2\pi\times0.3$ for the last one. 


Figures \ref{Fig4} (b1)-(b3) show $E(t)$, $\delta\Delta_{1,2}'(t)$, and $\delta\varphi(t)$, respectively, for $\Omega=0.25$, $\tau=2\pi\times1$. 
In this case, photon energy $\Omega<2\Delta_1$ is insufficient to excite Bogoliubov quasiparticles. 
Also, it deviates from the resonance condition for the Higgs mode ($2\Omega\simeq2\Delta_{1,2}$). 
As a result, the gap function varies only through virtual excitation of quasiparticles 
and approaches 0 right after the electric field vanishes.


Figures \ref{Fig4} (c1)-(c3) show the case with $\Omega=0.5$. 
Photon energy is again insufficient to excite quasiparticles ($\Omega<2\Delta_1$), 
but now satisfies the resonance condition for the lower-energy Higgs mode, $2\Omega\simeq2\Delta_1$. 
As a result, a small free oscillation remains in both gaps after illumination. 


Figures \ref{Fig4} (d1)-(d3) show the case with $\Omega=0.75$. 
Still, quasiparticles are not excited because $\Omega<2\Delta_1$, and the resonance condition $2\Omega\simeq2\Delta_{1,2}$ is not met again. 
Correspondingly, the gap functions rapidly approach 0 after forced oscillation. 
In fact, $2\Omega=1.5$ is close to the resonance condition for the Leggett mode, because its energy is about 1.3 in the present model \cite{Murotani2017}. 
However, Fig. \ref{Fig4} (d3) shows no remarkable structure because it is damped. 


Figures \ref{Fig4} (e1)-(e3) show the case with $\Omega=1$. 
Now, the power spectrum overlaps with the onset of conductivity ($2\Delta_1$) as seen in Fig. \ref{Fig4} (a), so that quasiparticles are excited. 
In addition, the resonance condition for the larger-energy Higgs mode $2\Omega\simeq2\Delta_2$ is satisfied, so that the Higgs mode is also excited.
As a result, $\delta\Delta_{1,2}'(t)$ oscillates even after the excitation because of the induced Higgs mode, 
and approaches a negative value because of the excited quasiparticles.


Figures \ref{Fig4} (f1)-(f3) show the case with $\Omega=1.5$. 
Only quasiparticles are excited because the photon energy is apart from the resonance conditions for collective modes. 
Correspondingly, the gap functions show a forced oscillation only under the electric field, and quickly approach the final values.


Finally, we examine the impulsive stimulated Raman scattering of low-energy light by collective modes. 
To this end, we considered a pulse with $\Omega=3$, $\tau=2\pi\times0.3$. 
The rightmost curve in Fig. \ref{Fig4} (a) plots the corresponding power spectrum, 
showing a bandwidth comparable with the smaller gap 
and thus sufficient to excite the lower-energy Higgs mode with a Raman process. 
The simulated result is shown in Figs. \ref{Fig4} (g1)-(g3).
Now the oscillation of the incident pulse is so fast that the real parts $\delta\Delta_{1,2}'(t)$ cannot follow it. 
As a result, they almost monotonically decrease within the pulse duration. 
After excitation ($t/2\pi>4.5$), however, a small oscillation remains, 
which reflects the Higgs mode induced by the Raman process. 
But its oscillation is blurred by the large gap reduction by quasiparticle excitations. 
On the other hand, oscillation of the Leggett mode is more easily seen in Fig. \ref{Fig4} (c). 

In ordinary Raman experiments, the incident photon energy exceeds $\gamma$ ($\sim0.1~\mathrm{eV}$), 
which makes the intraband transitions assisted by the impurity scattering irrelevant. 
Then, electronic Raman scattering will be dominated by interband transitions \cite{Klein1984,Klein2010}.
This consideration may explain the reason why the Higgs mode in MgB$_2$ has not been observed through the spontaneous Raman scattering \cite{Blumberg2007}.

To be exact, relaxation processes will modify light-induced dynamics of the gap functions. 
For example, strong electron-phonon interaction gives rise to damping of the Higgs mode at high temperatures \cite{Murakami2016} 
while carrier-carrier scattering redistributes the photoexcited quasiparticles. 
However, it is reasonable to expect that the calculated results will not be significantly altered, especially just under illumination of the optical pulse. 
Thus we do not consider relaxation processes other than the impurity scattering already taken into account. 

\section{Third-harmonic generation}

\begin{figure}
\centering
\includegraphics[width=8.5cm]{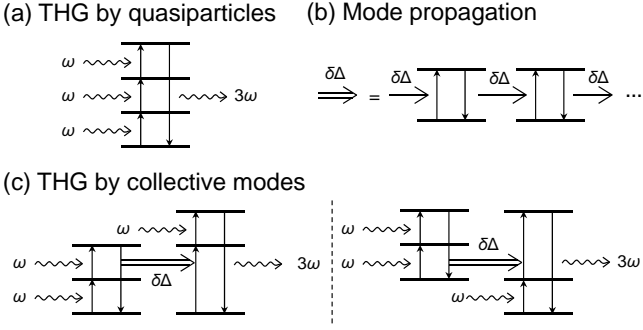}
\caption{Schematic level diagrams concerning THG in BCS superconductors. 
(a) Contribution from quasiparticle excitations. 
Horizontal lines denote quasiparticle levels, either real or virtual.
Wavy arrows indicate photons. 
(b) Propagation of collective modes. 
Successive absorption and emission of $\delta\Delta$ (single-line arrows) induce 
the self-consistent motion of collective modes (double-line arrows), either Higgs or Leggett.
(c) Contribution from collective modes. 
}\label{Fig5}
\end{figure}

Let us now consider THG. 
We start from a brief discussion of its origin. 
Figure \ref{Fig5} schematically shows the processes that concern THG in BCS superconductors. 
First, Fig. \ref{Fig5} (a) gives the simplest one in which a Bogoliubov quasiparticle absorbs three photons and emits one photon. 
No collective mode concerns this process. 
However, quasiparticles can emit not only photons but also collective modes. 
In this point of view, Eqs. (\ref{real_part2}) and (\ref{imaginary_part2}) can be regarded as the solution of a wave equation for $\delta\Delta$ with the source $r^{ab}(\epsilon)$. 
The emitted $\delta\Delta$ then formally acts as an external field in Eq. (\ref{dHint}) and thus is absorbed by quasiparticles again. 
Such a sequence induces the collective modes as depicted in Fig. \ref{Fig5} (b). 
As a result, the collective modes can also contribute to THG through diagrams shown in Fig. \ref{Fig5} (c).

Below, we assume that an isotropic impurity scattering mixes the states close to the Fermi energy almost equally. 
Then, direction of a quantum number $\kk$ can be neglected, which enables us to approximate $|\mathbf{e}\cdot\J_{i\kk\kk'}|^2\simeq\langle|\mathbf{e}\cdot\J_{i\kk\kk'}|^2\rangle_{\mathrm{Av}}$. 
Then, it follows that
\begin{align}
&\left\langle\left|\textbf{e}\cdot\J_{i\kk\kk'}\right|^2\left|\textbf{e}\cdot\J_{i\kk\kk''}\right|^2\right\rangle_{\mathrm{Av}}\nonumber\\
&\quad\simeq\left\langle\left|\textbf{e}\cdot\J_{i\kk\kk'}\right|^2\right\rangle_{\mathrm{Av}}\left\langle\left|\textbf{e}\cdot\J_{i\kk\kk''}\right|^2\right\rangle_{\mathrm{Av}}. 
\end{align}
In fact, the right-hand side gives the minimum among the possible values of the left-hand side.
The possible maximum is $9/5$ times larger, which is not far from the above equation.
Therefore, we expect that the exact form of the transition matrix elements does not matter significantly to the final results.

The third-order current density is given by 
\begin{align}
\jj(t)\Big|_3&=\jj_{\mathrm{P}}(t)\Big|_3+\jj_{\mathrm{D}}(t)\Big|_3,
\end{align}
where 
\begin{align}
\jj_{\mathrm{P}}(t)\Big|_3&=\mathbf{e}\sum_i\frac{e^2n_i}{2m_i}\iint d\epsilon~d\epsilon'~W_i(\epsilon,\epsilon')\nonumber\\
&\quad\times\left\{l_i(\epsilon,\epsilon')\left[R_i^{11}(\epsilon,\epsilon')+R_i^{22}(\epsilon,\epsilon')\right]\right.\nonumber\\
&\quad\quad\left.+p_i(\epsilon,\epsilon')\left[R_i^{21}(\epsilon,\epsilon')-R_i^{12}(\epsilon,\epsilon')\right]\right\},\label{jP3-1}\\
\jj_{\mathrm{D}}(t)\Big|_3&=-\sum_is_i\frac{e^2\A}{m_i}N_i(0)\int d\epsilon\nonumber\\
&\quad\times\left\{\left[u_i(\epsilon)^2-v_i(\epsilon)^2\right]\left[r_i^{11}(\epsilon)-r_i^{22}(\epsilon)\right]\right.\nonumber\\
&\quad\quad\left.+2u_i(\epsilon)v_i(\epsilon)\left[r_i^{21}(\epsilon)+r_i^{12}(\epsilon)\right]\right\},\label{jD3-1}
\end{align}
and 
\begin{align}
R_i^{ab}(\epsilon,\epsilon')=\frac{\left\langle\left(\mathbf{e}\cdot\J_{i\kk\kk'}\right)\rho_{i\kk\kk'}^{ab}\Big|_3\right\rangle_{\mathrm{Av}}}
{\left\langle\left|\textbf{e}\cdot\J_{i\kk\kk'}\right|^2\right\rangle_{\mathrm{Av}}}.
\end{align}
The function $R_i^{ab}(\epsilon,\epsilon')$ follows the equation of motion 
\begin{align}
&\left[i\frac{\partial}{\partial t}-(E'-E)\right]R_i^{11}(\epsilon,\epsilon')\nonumber\\
&\quad=l_i(\epsilon',\epsilon)\left\{A\left[r_i^{11}(\epsilon')-r_i^{11}(\epsilon)\right]\right.\nonumber\\
&\quad\quad\left.+\delta\Delta_i'\left(\frac{\Delta_i}{E'}-\frac{\Delta_i}{E}\right)F_i^{11}(\epsilon,\epsilon')\right\}\nonumber\\
&\quad\quad+p_i(\epsilon',\epsilon)\left\{A\left[r_i^{21,\mathrm{odd}}(\epsilon')+r_i^{21,\mathrm{odd}}(\epsilon)^*\right]\right.\nonumber\\
&\quad\quad\left.+\delta\Delta_i'\left(-\frac{\epsilon'}{E'}F_i^{12}(\epsilon,\epsilon')-\frac{\epsilon}{E}F_i^{21}(\epsilon,\epsilon')\right)\right\},\\
&\left[i\frac{\partial}{\partial t}-(E'+E)\right]R_i^{21}(\epsilon,\epsilon')\nonumber\\
&\quad=-p_i(\epsilon',\epsilon)\left\{A\left[r_i^{11}(\epsilon')+r_i^{11}(\epsilon)\right]\right.\nonumber\\
&\quad\quad\left.+\delta\Delta_i'\left(\frac{\Delta_i}{E'}+\frac{\Delta_i}{E}\right)F_i^{21}(\epsilon,\epsilon')\right\}\nonumber\\
&\quad\quad+l_i(\epsilon',\epsilon)\left\{A\left[r_i^{21,\mathrm{odd}}(\epsilon')-r_i^{21,\mathrm{odd}}(\epsilon)\right]\right.\nonumber\\
&\quad\quad\left.+\delta\Delta_i'\left(\frac{\epsilon}{E}F_i^{11}(\epsilon,\epsilon')-\frac{\epsilon'}{E'}F_i^{22}(\epsilon,\epsilon')\right)\right\},\\
&R_i^{22}(\epsilon,\epsilon')=R_i^{11}(\epsilon,\epsilon')^*,~R_i^{12}(\epsilon,\epsilon')=-R_i^{21}(\epsilon,\epsilon')^*,
\end{align}
under the particle-hole symmetry. 
Only the Higgs mode and quasipaticle excitations contribute to the paramagnetic component. 
An order estimation gives  
\begin{gather}
\jj_{\mathrm{P}}(t)\Big|_3\sim\frac{e^4n}{m^2\eF}A^3\times
\begin{cases}
(\eF\gamma/\Delta^2)^2 & \gamma\ll2\Delta, \\
(\eF/\gamma)^2 & \gamma\gg2\Delta.\label{HM_QP_Para}
\end{cases}\\
(\mathrm{HM\&QP/Para})\nonumber
\end{gather}
Here, HM and QP are abbreviations of the Higgs mode and quasiparticles, respectively.

When a small particle-hole asymmetry is taken into account, the Leggett mode (LM) also contributes to $\jj_{\mathrm{P}}(t)|_3$.
However, its order is small compared to the others:
\begin{gather}
\jj_{\mathrm{P}}(t)\Big|_3\sim\frac{e^4n}{m^2\eF}A^3\times
\begin{cases}
(\gamma/\Delta)^2 & \gamma\ll2\Delta, \\
(\Delta/\gamma)^2 & \gamma\gg2\Delta.
\end{cases}\\
(\mathrm{LM/Para})\nonumber
\end{gather}
Therefore, this contirbution is negligible. 

\begin{table}
\centering
\begin{tabular}{@{\hspace{3mm}}c@{\hspace{3mm}}||@{\hspace{3mm}}c@{\hspace{3mm}}|@{\hspace{3mm}}c@{\hspace{3mm}}} \hline
Channel & Para ($\pp\cdot\A$) & Dia ($\A^2$) \\ \hline
HM & $(\eF\gamma/\Delta^2)^2~\to~(\eF/\gamma)^2$ & 0 [$(\Delta/\eF)^2$] \\ 
QP & $(\eF\gamma/\Delta^2)^2~\to~(\eF/\gamma)^2$ & 0 [1] \\
LM & $(\gamma/\Delta)^2~\to~(\Delta/\gamma)^2~(*)$ & 1 \\ \hline
\end{tabular}
\caption{Order of the third-order current $\jj(t)|_3$ in the unit of $(e^4n/m^2\eF)A^3$. 
Arrows connects the clean limit ($\gamma\ll2\Delta$) on the left side and the dirty limit ($\gamma\gg2\Delta$) on the right side. 
For reference, square brackets show the case when the energy dispersion relation exhibits a nonparabolicity, calculated in the clean limit \cite{Tsuji2015,Murotani2017}. 
The asterisked part is neglected in the main text.
The combination of QP and Dia is specially called charge-density fluctuations.
}\label{TableII}
\end{table}

We turn to the diamagnetic component. 
For a parabolic and isotropic dispersion relation, only $r_i^{21,\mathrm{even}}(\epsilon)$ contributes to Eq. (\ref{jD3-1}). 
In other words, only the Leggett mode contributes to the diamagnetic component. 
This is understood in terms of Eq. (\ref{jD1-1}), which can be extended to nonlinear current by allowing carrier density $n_i$ to vary. 
While the Leggett mode changes $n_i$ through the interband Josephson coupling $U_{12}$, 
the Higgs mode and quasiparticles do not. 
Therefore, the latter two do not contribute to the diamagnetic component of the third-order current.
Its order is then estimated as
\begin{align}
\jj_{\mathrm{D}}(t)\Big|_3&\sim\frac{e^4n}{m^2\eF}A^3.\quad(\mathrm{LM/Dia})
\end{align}
In the clean limit, however, it is known that a nonparabolicity of the dispersion relation enables the Higgs mode and quasiparticles to induce nonzero $\jj_{\mathrm{D}}(t)$ \cite{Tsuji2015,Murotani2017}.
The corresponding correction is given by 
\begin{align}
\jj_{\mathrm{D}}(t)\Big|_3&\sim\frac{e^4n}{m^2\eF}A^3\times
\begin{cases}
(\Delta/\eF)^2, & (\mathrm{HM/Dia}) \\
1. & (\mathrm{QP/Dia})
\end{cases}\label{HM_QP_Dia}
\end{align}
Again, it is reasonable to assume the validity of this order estimation even for dirty cases, 
because of the insensitivity of the interaction Hamiltonian (\ref{H2}) against the impurity scattering. 
The order estimation is summarized in Table \ref{TableII}.

As already mentioned, only the diamagnetic coupling contributes to THG in the clean limit. 
And the phase degree of freedom is negligible in single-band systems. 
As a result, the combination of quasiparticles and the diamagnetic coupling (called charge-density fluctuations) 
gives the dominant origin of THG \cite{Cea2016-1}, unless one includes the retardation effect beyond the BCS mean-field treatment \cite{Tsuji2016}.
The Leggett mode also contributes in the same order in multiband systems \cite{Murotani2017}.
In dirty systems, however, Eq. (\ref{HM_QP_Dia}) is negligible compared to Eq. (\ref{HM_QP_Para}). 
Therefore, in most superconductors exhibiting a dirty nature ($2\Delta<\gamma$), 
the dominant origin of THG will be the Higgs mode and quasiparticles excited by the paramagnetic coupling.

The above order estimation cannot reveal the relative importance of the Higgs mode and quasiparticles. 
To examine it, we considered a two-band system with the same parameters as in Sec. \ref{SecIII}.
For simplicity, we concentrated on $T=0.02$ well below $\Tc=0.54$ 
and on a multicycle pulse (\ref{vector_potential}) with $\Omega=1$ and $\tau=2\pi\times1$, 
which was used also in Fig. \ref{Fig4} (e). 
The chosen frequency satisfies the resonance condition $2\Omega\simeq2\Delta_2$ 
both for the Higgs mode with the larger energy and for the quasiparticle excitations with the larger gap. 

\begin{figure}
\centering
\includegraphics[width=8.5cm]{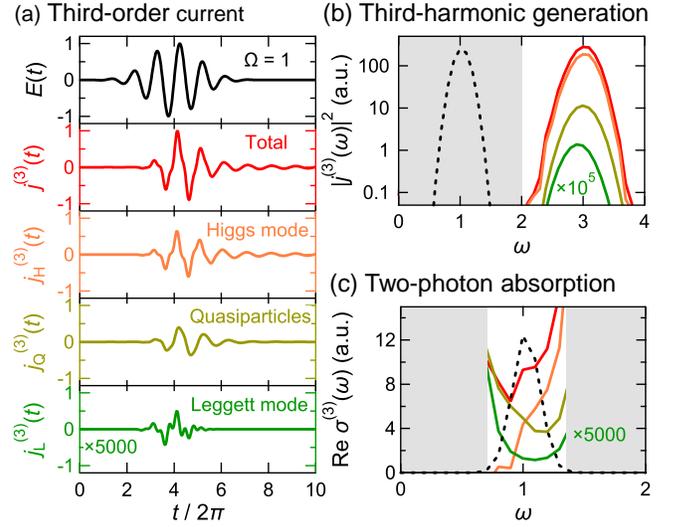}
\caption{Third-order nonlinear response of a two-band BCS superconductor. 
(a) Top: electric field waveform. 
Second: total third-order current. 
Third, fourth, bottom: contribution from the Higgs mode, quasiparticles, and the Leggett mode, respectively, 
with the last one multiplied by 5000.
(b) Intensity of the third harmonic. The topmost line gives the total emission. 
The other three lines show contributions from the Higgs mode, quasipartlces, and the Leggett mode, from top to bottom. 
The last one is multiplied by $10^5$. 
Dashed line is intensity of the incident electric field. a.u. stands for arbitrary units.
(c) Nonlinear absorption spectrum. 
The shaded region is outside the detection range of the incident electric field, 
whose intensity is shown as the dashed line. 
The topmost line gives the total spectrum. 
The other three lines are contributions from the Higgs mode, quasipartlces, and the Leggett mode, from top to bottom (seen at the edge of the right shaded region).
The last one is multiplied by 5000.
}\label{Fig6}
\end{figure}

The electric field waveform is plotted in the top panel of Fig. \ref{Fig6} (a) 
with the induced third-order current $j^{(3)}(t)=\mathbf{e}\cdot\jj(t)|_3$ in the next panel. 
We can decompose the latter into contributions from the Higgs mode, quasiparticles, and the Leggett mode, 
which are shown in the subsequent three panels in order.
In our approximations, the diamagnetic component $j^{(3)}_{\mathrm{L}}(t)=\mathbf{e}\cdot\jj_{\mathrm{D}}(t)|_3$ corresponds to the Leggett mode (green). 
This is about 5000 times smaller than the others, 
being consistent with the order estimation predicting a $(\eF/\gamma)^2\sim10^3$ times smaller contribution for the chosen parameters.
The contribution from quasiparticles $j^{(3)}_{\mathrm{Q}}(t)=\mathbf{e}\cdot\jj_{\mathrm{P}}(t)|_{3,\delta\Delta'=0}$ is obtained 
by neglecting the self-consistency condition (\ref{real_part2}), i.e., artificially putting $\delta\Delta'=0$ (yellow).
The remaining part $j^{(3)}_{\mathrm{H}}(t)=\mathbf{e}\cdot[\jj_{\mathrm{P}}(t)|_3-\jj_{\mathrm{P}}(t)|_{3,\delta\Delta'=0}]$ arises from the Higgs mode (orange). 

All these contributions consist of components with the frequency $\Omega+\Omega+\Omega=3\Omega$ (THG) and with $\Omega-\Omega+\Omega=\Omega$ (two-photon absorption).
First, we examine the former.
In Fig. \ref{Fig6} (b), we plot the power spectrum of the induced current $j^{(3)}(t)$ for $2\le\omega\le4$. 
The topmost (red) curve corresponds to the total third-order current, 
peaked at the third-harmonic frequency $\omega=3\Omega=3$. 
The lower three lines are given by considering only the Higgs mode, only quasiparticles, and only the Leggett mode, from top to bottom, respectively. 
As seen easily, the total third harmonic is dominated by the Higgs mode (orange), while
the contribution from quasiparticles is about one order of magnitude smaller (yellow). 
This behavior is similar to the conductivity spectrum of a single-band superconductor with a dc supercurrent 
featured by a modification by the Higgs mode larger than by quasiparticles \cite{Moor2017,Nakamura2018}. 
Because the contribution from the Leggett mode is so small that it is originally outside the plotted region, 
the corresponding curve is multiplied by $10^5$ (green). 
This numerical simulation thus reveals that the Higgs mode plays the dominant role in THG, 

Finally, we mention the two-photon absorption. 
We introduce a nonlinear absorption spectrum 
\begin{align}
\operatorname{Re}\sigma^{(3)}(\omega)=\operatorname{Re}\left[\frac{j^{(3)}(\omega)}{E(\omega)}\right].\label{sigma3}
\end{align}
This function gives the nonlinear correction to the absorption spectrum effectively felt by the incident electric field itself. 
We plotted the calculated spectra in Fig. \ref{Fig6} (c). 
Because Eq. (\ref{sigma3}) condenses all sum- and difference-frequency generation processes into dependence on a single frequency $\omega$, 
an unphysical upturn at the edges of the pump bandwidth appears. 
However, it does not matter because the absorbed energy is proportional to $\operatorname{Re}\sigma^{(3)}(\omega)$ times $|E(\omega)|^2$.
In Fig. \ref{Fig6} (c), it can be seen that contributions from the Higgs mode (orange) and quasiparticles (yellow) are comparable, 
while that from the Leggett mode (green) is small (it is shown after multiplication by 5000).
This observation confirms the result of Sec. \ref{SecIV} which found that two-photon excitation of the Higgs mode can occur more efficiently than the Leggett mode.

\section{Conclusion}

In summary, we investigated the nonlinear optical response of weak-coupling multiband superconductors containing nonmagnetic impurities with a density matrix approach. 
We found that impurity scattering enhances the light-induced Higgs mode through the paramagnetic coupling with the gauge field, 
while the Leggett mode is left hardly affected. 
Consequently, light-induced non-equilibrium dynamics of superconducting gaps is dominated by the Higgs mode for a low-energy excitation. 
We also studied THG in dirty multiband superconductors, 
revealing the dominant contribution from the Higgs mode. 
Contribution from quasiparticle excitations is smaller, 
and Leggett mode will be negligible in this phenomenon. 

It is an interesting problem to quantitatively compare the enhancement of the Higgs mode by nonmagnetic impurities 
with another enhancement by retarded interaction \cite{Tsuji2016}.
In addition, to the best of our knowledge, it is not known how the interband impurity scattering affects the resonance structure of collective modes. 
Use of Green's function method, which can take into account the retardation effect of phonon-mediated interaction \cite{Murakami2016,Tsuji2016}, 
interband impurity scattering \cite{Sung1967}, and also magnetic impurities \cite{Jujo2018} etc., may pave the way for more detailed understanding of non-equilibrium properties of superconductors.

\begin{acknowledgments}
We wish to thank D. Manske and K. Tomita for fruitful discussions. 
R.S. acknowledges partial support by JSPS KAKENHI Grant Nos. 18H05324 and 15H02102. 
Y.M. is supported by JSPS Research Fellowship for Young Scientists.
\end{acknowledgments}

\appendix

\section{Derivation of transition matrix element}\label{matrix_element}

In this section, we derive Eq. (\ref{MB1}) according to the method of MB \cite{Mattis1958}.
MB assumed a correlation function in a form of
\begin{align}
\rho_k(R)\equiv\left\langle\phi_{\kk}^*(\rr)\phi_{\kk}(\rr')\right\rangle_{\mathrm{Av}}=\frac{\sin kR}{kR}e^{-R/2l}, \label{correlation_function}
\end{align}
where $\phi_{\kk}(\rr)$ is an eigenfunction of the noninteracting Hamiltonian including impurity potential, 
$\langle~\rangle_{\mathrm{Av}}$ denotes an average over the angle of $\kk$, 
$\mathbf{R}=\rr-\rr'$, and $l$ is the mean free path. 
We assume the above equation for each band independently.
From definition, the transition matrix element is given by 
\begin{align}
\left|J_{\kk\kk'}^i\right|^2
&=\frac{e^2}{m^2}\left|\int d^3\rr~\phi_{\kk}^*(\rr)\frac{1}{i}\frac{\partial}{\partial x_i}\phi_{\kk'}(\rr)\right|^2, 
\end{align}
for $i=x,y,z$. 
Averaging over directions of $\kk$ and $\kk'$ yields
\begin{align}
\left\langle\left|J_{\kk\kk'}^i\right|^2\right\rangle_{\mathrm{Av}}
&=\frac{e^2}{m^2}\int d^3\mathbf{R}~\frac{\partial\rho_k(R)}{\partial R_i}\frac{\partial\rho_{k'}(R)}{\partial R_i}.
\end{align}
Note that we have set the volume of the system as unity.
Using Eq. (2.13) in MB, this integral is reduced to
\begin{align}
\left\langle\left|J_{\kk\kk'}^i\right|^2\right\rangle_{\mathrm{Av}}
&\simeq\frac{2\pi e^2}{3m^2}\frac{l^{-1}}{(k-k')^2+l^{-2}},\label{matrix_element_k}
\end{align}
for $k,k'\simeq\kF$. 
Substituting $\epsilon_{\kk}\simeq\vF(k-\kF)$ into Eq. (\ref{matrix_element_k}), we obtain Eq. (\ref{MB1}) with $\gamma=\vF/l$.

\section{Quasiparticle representation}\label{quasiparticle_representation}

In this section, we construct a density matrix method for Bogoliubov quasiparticles.
First, in the thermal equilibrium where the mean-field $\Delta_i$ is constant, 
the Hamiltonian can be diagonalized in a form of
\begin{align}
\mathcal{H}_0+\mathcal{H}_{\mathrm{int}}^{\mathrm{eq}}&=\sum_{i\kk}\psi_{i\kk}^\dag\left(
\begin{array}{cc}
E_{i\kk} & 0 \\
0 & -E_{i\kk}
\end{array}
\right)\psi_{i\kk},\label{Hdiag}
\end{align}
where we have performed Bogoliubov transformation
\begin{align}
\psi_{i\kk}=
\left(
\begin{array}{cc}
u_{i\kk} & -v_{i\kk} \\ v_{i\kk}^* & u_{i\kk}
\end{array}
\right)\left(
\begin{array}{c}
c_{i\kk\uparrow} \\ c_{i(-\kk)\downarrow}^\dag
\end{array}
\right),\label{spinor}
\end{align}
with
\begin{align}
E_{i\kk}&=\sqrt{\epsilon_{i\kk}^2+|\Delta_i^{\mathrm{eq}}|^2},~
u_{i\kk}^2=\frac{1}{2}\left(1+\frac{\epsilon_{i\kk}}{E_{i\kk}}\right),\nonumber\\
|v_{i\kk}|^2&=\frac{1}{2}\left(1-\frac{\epsilon_{i\kk}}{E_{i\kk}}\right),~
2u_{i\kk}v_{i\kk}=\frac{\Delta_i^{\mathrm{eq}}}{E_{i\kk}}.
\end{align}
Here, the superscript ``eq'' labels the equilibrium values. 
Assuming Fermi statistics of Bogoliubov quasiparticles, 
Eq. (\ref{self_consistency}) is explicitly rewritten as 
\begin{align}
\Delta_i^{\mathrm{eq}}
&=\sum_jU_{ij}N_{j}(0)\Delta_j^{\mathrm{eq}}\int_{-\wD}^{\wD}\frac{d\epsilon}{2\sqrt{\epsilon^2+(\Delta_j^{\mathrm{eq}})^2}}\nonumber\\
&\quad\times\tanh\left(\frac{\beta}{2}\sqrt{\epsilon^2+(\Delta_j^{\mathrm{eq}})^2}\right),\label{gap_eqn}
\end{align}
where $\wD$ is the Debye frequency. 
Equation (\ref{gap_eqn}) gives the gap equation for a multiband BCS superconductor. 
We will consider $U_{ij}$ to be a real number, which allows us to choose real values for $\Delta_i^{\mathrm{eq}}$ and $v_{i\kk}$ at the equilibrium. 

Next, we present the equation of motion for the density matrix (\ref{density_matrix}).
For that purpose, we construct a 4-component vector 
\begin{align}
\vec{\rho}_{i\kk\kk'}&=\left(
\begin{array}{cc}
\rho_{i\kk\kk'}^{11} \\ 
\rho_{i\kk\kk'}^{12} \\ 
\rho_{i\kk\kk'}^{21} \\
\rho_{i\kk\kk'}^{22}
\end{array}
\right). 
\end{align}
When the total Hamiltonian is expressed as
\begin{align}
\mathcal{H}&=\sum_{ab}\sum_{i\kk\kk'}\psi_{i\kk}^{a\dag}h_{i\kk\kk'}^{ab}\psi_{i\kk'}^b,
\end{align}
the equation of motion for the 4-component vector is given by 
\begin{align}
i\frac{\partial\vec{\rho}_{i\kk\kk'}}{\partial t}=\sum_{\kk''}
\left[H_{i\kk'\kk''}^{(1)}\vec{\rho}_{i\kk\kk''}-H_{i\kk''\kk}^{(2)}\vec{\rho}_{i\kk''\kk'}\right],\label{EoM}
\end{align}
with 
\begin{align}
H_{i\kk'\kk}^{(1)}&=\left(
\begin{array}{cccc}
h_{i\kk'\kk}^{11} & h_{i\kk'\kk}^{12} & 0 & 0 \\
h_{i\kk'\kk}^{21} & h_{i\kk'\kk}^{22} & 0 & 0 \\
0 & 0 & h_{i\kk'\kk}^{11} & h_{i\kk'\kk}^{12} \\
0 & 0 & h_{i\kk'\kk}^{21} & h_{i\kk'\kk}^{22} \\
\end{array}
\right),\nonumber\\
H_{i\kk'\kk}^{(2)}&=\left(
\begin{array}{cccc}
h_{i\kk'\kk}^{11} & 0 & h_{i\kk'\kk}^{21} & 0 \\
0 & h_{i\kk'\kk}^{11} & 0 & h_{i\kk'\kk}^{21} \\
h_{i\kk'\kk}^{12} & 0 & h_{i\kk'\kk}^{22} & 0 \\
0 & h_{i\kk'\kk}^{12} & 0 & h_{i\kk'\kk}^{22} \\
\end{array}
\right).
\end{align}
The equilibrium values are given by  
\begin{align}
\vec{\rho}_{i\kk\kk'}\Big|_0&=\delta_{\kk\kk'}\left(
\begin{array}{c}
f_{i\kk} \\ 0 \\ 0 \\ 1-f_{i\kk}
\end{array}
\right),~f_{i\kk}=\frac{1}{e^{\beta E_{i\kk}}+1},\label{initial_value}\\
H_{i\kk'\kk}^{(1)}\Big|_0&=\delta_{\kk'\kk}E_{i\kk}\left(
\begin{array}{cccc}
1 & 0 & 0 & 0 \\
0 & -1 & 0 & 0 \\
0 & 0 & 1 & 0 \\
0 & 0 & 0 & -1
\end{array}
\right),\\
H_{i\kk'\kk}^{(2)}\Big|_0&=\delta_{\kk'\kk}E_{i\kk}\left(
\begin{array}{cccc}
1 & 0 & 0 & 0 \\
0 & 1 & 0 & 0 \\
0 & 0 & -1 & 0 \\
0 & 0 & 0 & -1
\end{array}
\right).
\end{align}

In addition to the equilibrium Hamiltonian (\ref{Hdiag}), 
there are three non-equilibrium terms given below.
First, when the gap function $\Delta_i(t)\equiv\Delta_i^{\mathrm{eq}}+\delta\Delta_i(t)$ is in motion, 
$\mathcal{H}_{\mathrm{int}}$ produces a perturbation Hamiltonian
\begin{align}
\delta\mathcal{H}_{\mathrm{int}}&\equiv\mathcal{H}_{\mathrm{int}}-\mathcal{H}_{\mathrm{int}}^{\mathrm{eq}}\nonumber\\
&=\sum_{i\kk}\psi_{i\kk}^\dag\left[\delta\Delta_i'\left(
\begin{array}{cc}
2u_{i\kk}v_{i\kk} & -u_{i\kk}^2+v_{i\kk}^2 \\
-u_{i\kk}^2+v_{i\kk}^2 & -2u_{i\kk}v_{i\kk}
\end{array}
\right)\right.\nonumber\\
&\quad\left.+\delta\Delta_i''\left(
\begin{array}{cc}
0 & -i \\ i & 0
\end{array}
\right)\right]\psi_{i\kk},\label{dHint}
\end{align}
where $\delta\Delta_i'=\operatorname{Re}\delta\Delta_i$, $\delta\Delta_i''=\operatorname{Im}\delta\Delta_i$. 
Using the density matrix, the gap function (\ref{self_consistency}) is rewritten as 
\begin{align}
\Delta_i'&=\sum_jU_{ij}\sum_{\kk}\left[-u_{j\kk}v_{j\kk}\left(\rho_{j\kk\kk}^{11}-\rho_{j\kk\kk}^{22}\right)\right.\nonumber\\
&\quad\left.+\frac{1}{2}(u_{j\kk}^2-v_{j\kk}^2)(\rho_{j\kk\kk}^{21}+\rho_{j\kk\kk}^{12})\right], \label{real_part}\\
\Delta_i''&=\sum_jU_{ij}\sum_{\kk}\frac{1}{2i}(\rho_{j\kk\kk}^{21}-\rho_{j\kk\kk}^{12}). \label{imaginary_part}
\end{align}
Motion of the real and imaginary parts corresponds to the Higgs and Leggett modes, respectively.
The latter can be viewed as follows. When $\Delta''_j$ varies from 0, 
it induces motion of $\rho_{j\kk\kk}^{21}$ and $\rho_{j\kk\kk}^{12}$ due to the coupling Hamiltonian (\ref{dHint}).
It then changes the value of $\Delta_i''$ with $i\neq j$ through Eq. (\ref{imaginary_part}).
Such an interaction between imaginary parts of different bands induces the normal mode of the phase difference, i.e., the Leggett mode.

On the other hand, the light-matter interaction is rewritten as 
\begin{align}
\mathcal{H}_1&=-\sum_{i\kk\kk'}\J_{i\kk\kk'}\cdot\A~\psi_{i\kk}^\dag\left(
\begin{array}{cc}
l_{i\kk\kk'} & -p_{i\kk\kk'} \\
p_{i\kk\kk'} & l_{i\kk\kk'}
\end{array}
\right)\psi_{i\kk'},\label{H1_qp}\\
\mathcal{H}_2&=\sum_{i\kk}s_i\frac{e^2\A^2}{2m_i}\psi_{i\kk}^\dag\left(
\begin{array}{cc}
u_{i\kk}^2-v_{i\kk}^2 & 2u_{i\kk}v_{i\kk} \\
2u_{i\kk}v_{i\kk} & -u_{i\kk}^2+v_{i\kk}^2
\end{array}
\right)\psi_{i\kk},\label{H2_qp}
\end{align}
where 
\begin{align}
l_{i\kk\kk'}&=u_{i\kk}u_{i\kk'}+v_{i\kk}v_{i\kk'},~p_{i\kk\kk'}=v_{i\kk}u_{i\kk'}-u_{i\kk}v_{i\kk'},\label{coherence_factors}
\end{align}
are coherence factors \cite{Schrieffer1964}.  
We have used $\J_{i(-\kk')(-\kk)}=-\J_{i\kk\kk'}$ which characterizes the case II interaction in the classification by BCS \cite{BCS1957}.
Using the density matrix, the paramagnetic and diamagnetic components of electric current density are given by 
\begin{align}
\textbf{j}_{\mathrm{P}}&=\sum_{i\kk\kk'}\J_{i\kk\kk'}\left[l_{i\kk\kk'}\left(\rho_{i\kk\kk'}^{11}+\rho_{i\kk\kk'}^{22}\right)\right.\nonumber\\
&\quad\quad\quad\quad\quad\quad\quad\left.+p_{i\kk\kk'}\left(\rho_{i\kk\kk'}^{21}-\rho_{i\kk\kk'}^{12}\right)\right], \label{jP}\\
\textbf{j}_{\mathrm{D}}&=-\sum_{i\kk}s_i\frac{e^2\A}{m_i}\left[(u_{i\kk}^2-v_{i\kk}^2)\left(\rho_{i\kk\kk}^{11}-\rho_{i\kk\kk}^{22}\right)\right.\nonumber\\
&\quad\quad\quad\quad\quad\quad\quad\left.+2u_{i\kk}v_{i\kk}\left(\rho_{i\kk\kk}^{21}+\rho_{i\kk\kk}^{12}\right)\right]. \label{jD}
\end{align}

\section{Derivation of linear response}\label{linear_response}

In this section, we derive Eq. (\ref{jP1-2}).
The corresponding equation of motion to be solved is 
\begin{align}
i\frac{\partial}{\partial t}\vec{\rho}_{i\kk\kk'}\Big|_1&=\left(H_{i\kk'\kk'}^{(1)}\Big|_0-H_{i\kk\kk}^{(2)}\Big|_0\right)\vec{\rho}_{i\kk\kk'}\Big|_1\nonumber\\
&\quad+H_{i\kk'\kk}^{(1)}\Big|_1\vec{\rho}_{i\kk\kk}\Big|_0-H_{i\kk'\kk}^{(2)}\Big|_1\vec{\rho}_{i\kk'\kk'}\Big|_0, \label{EoM1}
\end{align}
where 
\begin{align}
H_{i\kk'\kk}^{(1)}\Big|_1&=-\J_{i\kk'\kk}\cdot\A\left(
\begin{array}{cccc}
l_{i\kk'\kk} & -p_{i\kk'\kk} & 0 & 0 \\
p_{i\kk'\kk} & l_{i\kk'\kk} & 0 & 0 \\
0 & 0 & l_{i\kk'\kk} & -p_{i\kk'\kk} \\
0 & 0 & p_{i\kk'\kk} & l_{i\kk'\kk}
\end{array}
\right),\nonumber\\
H_{i\kk'\kk}^{(2)}\Big|_1&=-\J_{i\kk'\kk}\cdot\A\left(
\begin{array}{cccc}
l_{i\kk'\kk} & 0 & p_{i\kk'\kk} & 0 \\
0 & l_{i\kk'\kk} & 0 & p_{i\kk'\kk} \\
-p_{i\kk'\kk} & 0 & l_{i\kk'\kk} & 0 \\
0 & -p_{i\kk'\kk} & 0 & l_{i\kk'\kk}
\end{array}
\right).
\end{align}
Equation (\ref{EoM1}) is solved in a form of 
\begin{align}
\vec{\rho}_{i\kk\kk'}\Big|_1&=\J_{i\kk'\kk}\cdot\mathbf{e}\left(
\begin{array}{cccc}
l_{i\kk'\kk} & 0 & 0 & 0 \\
0 & p_{i\kk'\kk} & 0 & 0 \\
0 & 0 & -p_{i\kk'\kk} & 0 \\
0 & 0 & 0 & l_{i\kk'\kk}
\end{array}
\right)\nonumber\\
&\quad\times\left(
\begin{array}{c}
F_{i}^{11}(\epsilon_{i\kk},\epsilon_{i\kk'}) \\
F_{i}^{12}(\epsilon_{i\kk},\epsilon_{i\kk'}) \\
F_{i}^{21}(\epsilon_{i\kk},\epsilon_{i\kk'}) \\
F_{i}^{22}(\epsilon_{i\kk},\epsilon_{i\kk'})
\end{array}
\right), \label{linear_solution}
\end{align}
where the function $F_i^{ab}(\epsilon,\epsilon')$ follows Eqs. (\ref{EoM1-1})-(\ref{EoM1-3}).
We can confirm that the gap function does not respond linearly to vector potential, i.e., 
\begin{align}
\delta\Delta_i\Big|_1=0,
\end{align}
by substituting Eq. (\ref{linear_solution}) into Eqs. (\ref{real_part}) and (\ref{imaginary_part}).
Because we have considered an isotropic system, the induced current is parallel to $\A$. 
As a result, the paramagnetic component of the current becomes 
\begin{align}
\jj_{\mathrm{P}}(t)\Big|_1&=2\mathbf{e}\sum_{i\kk\kk'}\left|\mathbf{e}\cdot\J_{i\kk\kk'}\right|^2\left[l_{i\kk\kk'}^2\operatorname{Re}F_i^{11}(\epsilon_{i\kk},\epsilon_{i\kk'})\right.\nonumber\\
&\quad\left.+p_{i\kk\kk'}^2\operatorname{Re}F_i^{21}(\epsilon_{i\kk},\epsilon_{i\kk'})\right].
\end{align}
Now, we replace the summation over $\kk$ by integration over energy, following 
\begin{align}
\sum_{\kk}\to\int\frac{d^3\kk}{(2\pi)^3}=N_i(0)\int_{-\infty}^\infty d\epsilon_{i\kk}\int\frac{d\Omega_{\kk}}{4\pi},
\end{align}
where $\Omega_{\kk}$ denotes the solid angle of $\kk$. 
Extension of the integration interval to $(-\infty,\infty)$ is justified because large $|\epsilon_{i\kk}|$ contributes only negligibly to the integral.
With the above replacement, we obtain
\begin{align}
&\jj_{\mathrm{P}}(t)\Big|_1=2\mathbf{e}\sum_{i}N_i(0)^2\iint d\epsilon~d\epsilon'\left\langle\left|\mathbf{e}\cdot\J_{i\kk\kk'}\right|^2\right\rangle_{\mathrm{Av}}\nonumber\\
&\times\left[l_{i}(\epsilon,\epsilon')^2\operatorname{Re}F_i^{11}(\epsilon,\epsilon')+p_{i}(\epsilon,\epsilon')^2\operatorname{Re}F_i^{21}(\epsilon,\epsilon')\right]. \label{jP1-1}
\end{align}
Substitution of Eq. (\ref{MB1}) into Eq. (\ref{jP1-1}) yields Eq. (\ref{jP1-2}) in the main text. 

As for the diamagnetic component, it is more convenient to go back to the interaction Hamiltonian (\ref{H2}) than to use expression (\ref{jD}). 
From Eq. (\ref{H2}), we obtain 
\begin{align}
\jj_{\mathrm{D}}(t)\Big|_1&
=-\A\sum_{i}\frac{e^2}{m_i}s_i\sum_{\kk\sigma}\langle c_{i\kk\sigma}^\dag c_{i\kk\sigma}\rangle\Big|_0. 
\end{align}
For an electron band ($s_i=+1$), this equation is reduced to Eq. (\ref{jD1-1}).
Because Eq. (\ref{jD1-1}) does not depend on the sign of $e$, it is also applicable to a hole band. 
Strictly speaking, one has to take into account the nonparabolicity in the dispersion relation far below the Fermi surface to derive Eq. (\ref{jD1-1}) for a hole band. 
This follows from a more general expression of the diamagnetic component of intraband current, 
\begin{align}
\jj_{\mathrm{D}}(t)&=-e^2\sum_{\kk}\left[(\A\cdot\nabla_{\kk})\nabla_{\kk}\epsilon_{\kk}\right]n_{\kk}\nonumber\\
&=e^2\sum_{\kk}\left[(\A\cdot\nabla_{\kk})\nabla_{\kk}\epsilon_{\kk}\right](1-n_{\kk}),
\end{align}
where the band and spin indices are omitted for simplicity. 
$n_{\kk}$ is the occupation number of an electron in a pure Bloch state (and thus is not $\langle c_{\kk}^\dag c_{\kk}\rangle$ in the main text).
The second line follows from $\nabla_{\kk}\epsilon_{\kk}=0$ at the edges of the Brillouin zone.
For a hole band, we have assumed Eq. (\ref{ek}) with $s=-1$ in the region where $1-n_{\kk}\neq0$.
Then, Eq. (\ref{jD1-1}) follows also for a hole band.

In numerical calculations, we introduce a cutoff in the energy integrals in Eq. (\ref{jP1-2}). 
Then, the real part of the optical conductivity $\sigma_1(\omega)$ is incorrectly given at a large photon energy $\omega$ beyond the cutoff. 
This leads to incorrect evaluation of the imaginary part $\sigma_2(\omega)$ even for a low photon energy because of the Kramers-Kronig relation. 
To improve the evaluation, it is convenient to follow MB's method \cite{Mattis1958}, rewriting 
\begin{align}
\jj_{\mathrm{D}}(t)\Big|_1&=\A\sum_{i}\frac{e^2n_i}{m_i}\iint d\epsilon~d\epsilon'\frac{f(\epsilon)-f(\epsilon')}{\epsilon-\epsilon'}W_i(\epsilon,\epsilon'), \label{jD1-2}
\end{align}
where $f(\epsilon)=1/(e^{\beta\epsilon}+1)$. 
Equation (\ref{jD1-2}) is identical to Eq. (\ref{jD1-1}) as long as the integration interval ranges infinity. 
Even when a cutoff is introduced, this expression guarantees $\sigma_2(\omega\to0)=0$ for the normal state above the critical temperature.

\section{Second-order equation of motion}\label{2nd-order}

In this section, we give the equation of motion which describes the second-order response of the system.
The relevant equation of motion is 
\begin{align}
i\frac{\partial}{\partial t}\vec{\rho}_{i\kk\kk}\Big|_2&=\left(H_{i\kk\kk}^{(1)}\Big|_0-H_{i\kk\kk}^{(2)}\Big|_0\right)\vec{\rho}_{i\kk\kk}\Big|_2\nonumber\\
&\quad+\sum_{\kk'}\left(H_{i\kk\kk'}^{(1)}\Big|_1\vec{\rho}_{i\kk\kk'}\Big|_1-H_{i\kk'\kk}^{(2)}\Big|_1\vec{\rho}_{i\kk'\kk}\Big|_1\right)\nonumber\\
&\quad+\left(H_{i\kk\kk}^{(1)}\Big|_2-H_{i\kk\kk}^{(2)}\Big|_2\right)\vec{\rho}_{i\kk\kk}\Big|_0. \label{EoM2}
\end{align}
The second-order Hamiltonian can be decomposed into contributions from the diamagnetic coupling (D), the Higgs mode (H), and the Leggett mode (L)  as 
\begin{align}
H_{i\kk\kk}^{(1,2)}\Big|_2&
=H_{i\kk\kk}^{(1,2)}\Big|_{2,\mathrm{D}}
+H_{i\kk\kk}^{(1,2)}\Big|_{2,\mathrm{H}}
+H_{i\kk\kk}^{(1,2)}\Big|_{2,\mathrm{L}}.\label{H_decomposition}
\end{align}
First, the diamagnetic contribution is 
\begin{align}
H_{i\kk\kk}^{(1)}\Big|_{2,\mathrm{D}}&=\frac{e^2\A^2}{2m_i}\frac{s_i}{E_{i\kk}}\left(
\begin{array}{cccc}
\epsilon_{i\kk} & \Delta_i & 0 & 0 \\
\Delta_i & -\epsilon_{i\kk} & 0 & 0 \\
0 & 0 & \epsilon_{i\kk} & \Delta_i \\
0 & 0 & \Delta_i & -\epsilon_{i\kk} \\
\end{array}
\right),\\
H_{i\kk\kk}^{(2)}\Big|_{2,\mathrm{D}}&=\frac{e^2\A^2}{2m_i}\frac{s_i}{E_{i\kk}}\left(
\begin{array}{cccc}
\epsilon_{i\kk} & 0  & \Delta_i  & 0 \\
0 & \epsilon_{i\kk} & 0 & \Delta_i \\
\Delta_i & 0 & -\epsilon_{i\kk} & 0 \\
0 & \Delta_i & 0 & -\epsilon_{i\kk} \\
\end{array}
\right). 
\end{align}
Second, the Higgs mode contribution is 
\begin{align}
H_{i\kk\kk}^{(1)}\Big|_{2,\mathrm{H}}&=\frac{\delta\Delta_i'|_2}{E_{i\kk}}\left(
\begin{array}{cccc}
\Delta_i & -\epsilon_{i\kk} & 0 & 0 \\
-\epsilon_{i\kk} & -\Delta_i & 0 & 0 \\
0 & 0 & \Delta_i & -\epsilon_{i\kk} \\
0 & 0 & -\epsilon_{i\kk} & -\Delta_i \\
\end{array}
\right),\\
H_{i\kk\kk}^{(2)}\Big|_{2,\mathrm{H}}&=\frac{\delta\Delta_i'|_2}{E_{i\kk}}\left(
\begin{array}{cccc}
\Delta_i & 0  & -\epsilon_{i\kk}  & 0 \\
0 & \Delta_i & 0 & -\epsilon_{i\kk} \\
-\epsilon_{i\kk} & 0 & -\Delta_i & 0 \\
0 & -\epsilon_{i\kk} & 0 & -\Delta_i \\
\end{array}
\right).
\end{align}
Finally, the Leggett mode contribution is 
\begin{align}
H_{i\kk\kk}^{(1)}\Big|_{2,\mathrm{L}}&=\delta\Delta_i''\Big|_2\left(
\begin{array}{cccc}
0 & -i & 0 & 0 \\
i & 0 & 0 & 0 \\
0 & 0 & 0 & -i \\
0 & 0 & i & 0
\end{array}
\right),\\
H_{i\kk\kk}^{(2)}\Big|_{2,\mathrm{L}}&=\delta\Delta_i''\Big|_2\left(
\begin{array}{cccc}
0 & 0 & i & 0 \\
0 & 0 & 0 & i \\
-i & 0 & 0 & 0 \\
0 & -i & 0 & 0
\end{array}
\right).
\end{align}
A straightforward calculation transforms the equation of motion (\ref{EoM2}) into Eqs. (\ref{EoM2-1}), (\ref{EoM2-2}), and (\ref{EoM2-3}),
for the angle-averaged density matrix $r_i^{ab}(\epsilon)$.

\section{Order estimation}\label{order_estimation}

In this section, we outline the order estimation of light-induced quantities.
We exemplify Eq. (\ref{HM_Para}) for $\delta\Delta'$ and Eq. (\ref{LM_Para}) for $\delta\Delta''$.
For simplicity, the absolute zero will be assumed below so that $F^{11}=0$.
When the incident photon energy is in the order of $\Delta$, we can approximate $i\partial/\partial t\sim\Delta$ for the order estimation.
This transforms Eqs. (\ref{EoM1-2}), (\ref{EoM2-1}), (\ref{EoM2-2}) into 
\begin{align}
\Delta F^{21}&\sim A,~\Delta r^{11}\sim S^{11},~\Delta r^{21,\mathrm{odd}}\sim S^{21}, 
\end{align}
where we have also used $E\sim\Delta$, $u^2\sim1$, etc. 
In the last, we have assumed $\delta\Delta'\sim S^{21}$ to be confirmed later. 
We now evaluate the order of $S^{ab}$ in Eqs. (\ref{S11}), (\ref{S21}).
In the clean limit, $W(\epsilon,\epsilon')$ in the integrand restricts the integral region to $|\epsilon-\epsilon'|\lesssim\gamma\ll2\Delta$.
In this case, we have 
\begin{align}
\int d\epsilon'~W(\epsilon,\epsilon')&\sim1,~
p(\epsilon,\epsilon')^2\sim l(\epsilon,\epsilon')p(\epsilon,\epsilon')\sim\frac{\gamma}{\Delta},
\end{align}
which yields
\begin{align}
S^{ab}\sim\frac{e^2A^2}{2m}\frac{\gamma\eF}{\Delta^2}\quad\Rightarrow\quad
\delta\Delta'\sim\frac{e^2A^2}{2m}\frac{\gamma\eF}{\Delta^2}.
\end{align}
The arrow follows from Eq. (\ref{real_part2}) putting $UN(0)\sim1$ and $\int d\epsilon\sim\Delta$. 
The above result confirms $\delta\Delta'\sim S^{21}$
and gives the upper part of Eq. (\ref{HM_Para}).
On the other hand, in the dirty limit, $W(\epsilon,\epsilon')\sim\gamma^{-1}$ is almost constant in the relevant energy scale.
We can then estimate 
\begin{align}
\int d\epsilon'~W(\epsilon,\epsilon')&\sim\frac{\Delta}{\gamma},~
p(\epsilon,\epsilon')^2\sim l(\epsilon,\epsilon')p(\epsilon,\epsilon')\sim1,
\end{align}
so that
\begin{align}
S^{ab}\sim\frac{e^2A^2}{2m}\frac{\eF}{\gamma}\quad\Rightarrow\quad
\delta\Delta'\sim\frac{e^2A^2}{2m}\frac{\eF}{\gamma},
\end{align}
which gives the lower part of Eq. (\ref{HM_Para}). 

To estimate the magnitude of the Leggett mode excited by the paramagnetic coupling, 
we have to take particle-hole asymmetry into account.
For example, particle-hole asymmetry induces a small even component of $S^{21}$ 
which should appear in the right-hand side of Eq. (\ref{EoM2-3}) as a source term.
The order of the even component is about $N'(0)\Delta/N(0)\sim\Delta/\eF$ times that of the odd component. 
This observation enables us to conclude that the magnitude of $\delta\Delta''$ induced by the paramagnetic coupling is 
$\eF/\Delta$ times smaller than that of $\delta\Delta'$ estimated above.
Then Eq. (\ref{LM_Para}) follows.

Order of the third-order current can be estimated in the same way.


\begin{thebibliography}{99}

\bibitem{Matsunaga2017-1}
	R. Matsunaga and R. Shimano, ``Nonlinear terahertz spectroscopy of Higgs mode in s-wave superconductors,'' Phys. Scr. \textbf{92}, 024003 (2017).

\bibitem{Volkov1973}
	A. F. Volkov and S. M. Kogan, Zh. Eksp. Teor. Fiz. \textbf{65}, 2038 (1973) 
[``Collisionless relaxation of the energy gap in superconductors,'' Sov. Phys. JETP \textbf{38}, 1018 (1974)].

\bibitem{Nambu1960}
	Y. Nambu, ``Quasi-particles and gauge invariance in the theory of superconductivity,'' Phys. Rev. \textbf{117}, 648 (1960).

\bibitem{Goldstone1961}
	J. Goldstone, ``Field theories with `superconductor' solutions,'' Nuovo Cimento \textbf{19}, 154 (1961).

\bibitem{Goldstone1962}
	J. Goldstone, A. Salam, and S. Weinberg, ``Broken symmetries,'' Phys. Rev. \textbf{127}, 965 (1962).

\bibitem{Anderson1963}
	P. W. Anderson, ``Plasmons, gauge invariance, and mass,'' Phys. Rev. \textbf{130}, 439 (1963).

\bibitem{Higgs1964-1}
	P. W. Higgs, ``Broken symmetries, massless particles and gauge fields,'' Phys. Lett. \textbf{12}, 132 (1964).

\bibitem{Englert1964}
	F. Englert and R. Brout, ``Broken symmetry and the mass of gauge vector mesons,'' Phys. Rev. Lett. \textbf{13}, 321 (1964).

\bibitem{Higgs1964-2}
	P. W. Higgs, ``Broken symmetries and the masses of gauge bosons,'' Phys. Rev. Lett. \textbf{13}, 508 (1964).

\bibitem{Guralnik1964}
	G. S. Guralnik, C. R. Hagen and T. W. B. Kibble, ``Global conservation laws and massless particles,'' Phys. Rev. Lett. \textbf{13}, 585 (1964).

\bibitem{Varma2002}
	C. M. Varma, ``Higgs boson in superconductors,'' J. Low Temp. Phys. \textbf{126}, 901 (2002).

\bibitem{Pekker2015}
	D. Pekker and C. M. Varma, ``Amplitude/Higgs modes in condensed matter physics,'' Annu. Rev. Condens. Matter Phys. \textbf{6}, 269 (2015).

\bibitem{Sooryakumar1980}
	R. Sooryakumar and M. V. Klein, ``Raman scattering by superconducting-gap excitations and their coupling to charge-density waves,'' Phys. Rev. Lett. \textbf{45}, 660 (1980).

\bibitem{Littlewood1981}
	P. B. Littlewood and C. M. Varma, ``Gauge-invariant theory of the dynamical interaction of charge density waves and superconductivity,'' Phys. Rev. Lett. \textbf{47}, 811 (1981).

\bibitem{Measson2014}
	M.-A. M\'{e}asson, Y. Gallais, M. Cazayous, B. Clair, P. Rodi\`{e}re, L. Cario and A. Sacuto, ``Amplitude Higgs mode in the 2$H$-NbSe$_2$ superconductor,'' Phys. Rev. B \textbf{89}, 060503(R) (2014).

\bibitem{Matsunaga2013}
	R. Matsunaga, Y. I. Hamada, K. Makise, Y. Uzawa, H. Terai, Z. Wang, and R. Shimano, ``Higgs amplitude mode in the BCS superconductors Nb$_{1-x}$Ti$_x$N induced by terahertz pulse excitation,'' Phys. Rev. Lett. \textbf{111}, 057002 (2013).

\bibitem{Matsunaga2014}
	R. Matsunaga, N. Tsuji, H. Fujita, A. Sugioka, K. Makise, Y. Uzawa, H. Terai, Z. Wang, H. Aoki, and R. Shimano, ``Light-induced collective pseudospin precession resonating with Higgs mode in a superconductor,'' Science \textbf{345}, 1145 (2014).

\bibitem{Matsunaga2017-2}
	R. Matsunaga, N. Tsuji, K. Makise, H. Terai, H. Aoki, and R. Shimano, ``Polarization-resolved terahertz third-harmonic generation in a single-crystal superconductor NbN: Dominance of the Higgs mode beyond the BCS approximation,'' Phys. Rev. B \textbf{96}, 020505(R) (2017).

\bibitem{Katsumi2018}
	K. Katsumi, N. Tsuji, Y. I. Hamada, R. Matsunaga, J. Schneeloch, R. D. Zhong, G. D. Gu, H. Aoki, Y. Gallais, and Ryo Shimano, ``Higgs mode in the $d$-wave superconductor Bi$_2$Sr$_2$CaCu$_2$O$_{8+x}$ driven by an intense terahertz pulse,'' Phys. Rev. Lett. \textbf{120}, 117001 (2018).

\bibitem{Chu2019}
	H. Chu, M.-J. Kim, K. Katsumi, S. Kovalev, R. D. Dawson, L. Schwarz, N. Yoshikawa, G. Kim, D. Putzky, Z. Z. Li, H. Raffy, S. Germanskiy, J.-C. Deinert, N. Awari, I. Ilyakov, B. Green, M. Chen, M. Bawatna, G. Christiani, G. Logvenov, Y. Gallais, A. V. Boris, B. Keimer, A. Schnyder, D. Manske, M. Gensch, Z. Wang, R. Shimano, and S. Kaiser, ``New collective mode in superconducting cuprates uncovered by Higgs spectroscopy,'' arXiv: 1901.06675.

\bibitem{Leggett1966}
	A. J. Leggett, ``Number-phase fluctuations in two-band superconductors,'' Prog. Theor. Phys. \textbf{36}, 901 (1966).

\bibitem{Sharapov2002}
	S. G. Sharapov, V. P. Gusynin and H. Beck, ``Effective action approach to the Leggett's mode in two-band superconductors,'' Eur. Phys. J. B \textbf{30}, 45 (2002).

\bibitem{Burnell2010}
	F. J. Burnell, J. Hu, M. M. Parish and B. A. Bernevig, ``Leggett mode in a strong-coupling model of iron arsenide superconductors,'' Phys. Rev. B \textbf{82}, 144506 (2010).

\bibitem{Blumberg2007}
	G. Blumberg, A. Mialitsin, B. S. Dennis, M. V. Klein, N. D. Zhigadlo, and J. Karpinski, ``Observation of Leggett’s collective mode in a multiband MgB$_2$ superconductor,'' Phys. Rev. Lett. \textbf{99}, 227002 (2007).

\bibitem{Klein2010}
	M. V. Klein, ``Theory of Raman scattering from Leggett's collective mode in a multiband superconductor: application to MgB$_2$,'' Phys. Rev. B \textbf{82}, 014507 (2010).

\bibitem{Cea2016-2}
	T. Cea and L. Benfatto, ``Signature of the Leggett mode in the $A_{1g}$ Raman response: From MgB$_2$ to iron-based superconductors,'' Phys. Rev. B \textbf{94}, 064512 (2016).

\bibitem{Murotani2017}
	Y. Murotani, N. Tsuji, and H. Aoki, ``Theory of light-induced resonances with collective Higgs and Leggett modes in multiband superconductors,'' Phys. Rev. B \textbf{95}, 104503 (2017).

\bibitem{Krull2016}
	H. Krull, N. Bittner, G. S. Uhrig, D. Manske, and A. P. Schnyder, ``Coupling of Higgs and Leggett modes in non-equilibrium superconductors,'' Nat. Commun. \textbf{7}, 11921 (2016).

\bibitem{Giorgianni2019}
	F. Giorgianni, T. Cea, C. Vicario, C. P. Hauri, W. K. Withanage, X. Xi, and L. Benfatto, ``Leggett mode controlled by light pulses,'' Nat. Phys. \textbf{15}, 341 (2019). 

\bibitem{Jujo2015}
	T. Jujo, ``Two-photon absorption by impurity scattering and amplitude mode in conventional superconductors,'' J. Phys. Soc. Jpn. \textbf{84}, 114711 (2015).


\bibitem{Jujo2018}
	T. Jujo, ``Quasiclassical theory on third-harmonic generation in conventional superconductors with paramagnetic impurities,'' J. Phys. Soc. Jpn. \textbf{87}, 024704 (2018).

\bibitem{Silaev2019}
	M. Silaev, ``Nonlinear electromagnetic response and Higgs mode excitation in BCS superconductors with impurities,'' arXiv: 1902.01666.

\bibitem{Ortolani2008}
	M. Ortolani, P. Dore, D. Di Castro, A. Perucchi, S. Lupi, V. Ferrando, M. Putti, I. Pallecchi, C. Ferdeghini, and X. X. Xi, ``Two-band parallel conductivity at terahertz frequencies in the superconducting state of MgB$_2$,'' Phys. Rev. B \textbf{77}, 100507(R) (2008).

\bibitem{Pimenov2013}
	A. Pimenov, S. Engelbrecht, A. M. Shuvaev, B. B. Jin, P. H. Wu, B. Xu, L. X. Cao, and E. Schachinger, ``Terahertz conductivity in FeSe$_{0.5}$Te$_{0.5}$ superconducting films,'' New J. Phys. \textbf{15}, 013032 (2013).

\bibitem{Tsuji2015}
	N. Tsuji and H. Aoki, ``Theory of Anderson pseudospin resonance with Higgs mode in superconductors,'' Phys. Rev. B \textbf{92}, 064508 (2015). 

\bibitem{Cea2016-1}
	T. Cea, C. Castellani and L. Benfatto, ``Non-linear optical effects and third-harmonic generation in superconductors: Cooper-pairs vs Higgs mode contribution,'' Phys. Rev. B \textbf{93}, 180507(R) (2016).

\bibitem{Cea2018}
	T. Cea, P. Barone, C. Castellani, and L. Benfatto, ``Polarization dependence of the third-harmonic generation in multiband superconductors,'' Phys. Rev. B \textbf{97}, 094516 (2018).

\bibitem{Mattis1958}
	D. C. Mattis and J. Bardeen, ``Theory of the anomalous skin effect in normal and superconducting metals,'' Phys. Rev. \textbf{111}, 412 (1958).

\bibitem{Zimmermann1991}
	W. Zimmermann, E. H. Brandt, M. Bauer, E. Seider, and L. Genzel, ``Optical conductivity of BCS superconductors with arbitrary purity,'' Physica C \textbf{183}, 99 (1991).

\bibitem{Berlinsky1993}
	A. J. Berlinsky, C. Kallin, G. Rose, and A.-C. Shi, ``Two-fluid interpretation of the conductivity of clean BCS superconductors,'' Phys. Rev. B \textbf{48}, 4074 (1993).

\bibitem{Suhl1959}
	H. Suhl, B. T. Matthias, and L. R. Walker, ``Bardeen-Cooper-Schrieffer theory of supeconductivity in the case of overlapping bands,'' Phys. Rev. Lett. \textbf{3}, 552 (1959).

\bibitem{Anderson1959}
	P. W. Anderson, ``Theory of dirty superconductors,'' J. Phys. Chem. Solids \textbf{11}, 26 (1959).

\bibitem{Stanev2014}
	V. Stanev and A. E. Koshelev, ``Complex state induced by impurities in multiband superconductors,'' Phys. Rev. B \textbf{89}, 100505(R) (2014).

\bibitem{Golubov1997}
	A. A. Golubov and I. I. Mazin, ``Effect of magnetic and nonmagnetic impurities on highly anisotropic superconductivity,'' Phys. Rev. B \textbf{55}, 15146 (1997).

\bibitem{Mazin2002}
	I. I. Mazin, O. K. Andersen, O. Jepsen, O. V. Dolgov, J. Kortus, A. A. Golubov, A. B. Kuz'menko, and D. van der Marel, ``Superconductivity in MgB$_2$: Clean or dirty?'' Phys. Rev. Lett. \textbf{89}, 107002 (2002).

\bibitem{Urata2016}
	T. Urata, Y. Tanabe, K. K. Huynh, Y. Yamakawa, H. Kontani, and K. Tanigaki, ``Superconductivity pairing mechanism from cobalt impurity doping in FeSe: Spin ($s_\pm$) or orbital ($s_{++}$) fluctuation,'' Phys. Rev. B \textbf{93}, 014507 (2016).

\bibitem{Yu2017}
	T. Yu and M. W. Wu, ``Gauge-invariant theory of quasiparticle and condensate dynamics in response to terahertz optical pulses in superconducting semiconductor quantum wells. I. $s$-wave superconductivity in the weak spin-orbit coupling limit,'' Phys. Rev. B \textbf{96}, 155311 (2017).

\bibitem{Yang2018a}
	F. Yang and M. W. Wu, ``Gauge-invariant microscopic kinetic theory of superconductivity in response to electromagnetic fields,'' Phys. Rev. B \textbf{98}, 094507 (2018).

\bibitem{Yang2018b}
	F. Yang and M. W. Wu, ``Gauge-invariant microscopic kinetic theory of superconductivity: application to electromagnetic response of Nambu-Goldstone and Higgs modes,'' arXiv: 1812.06622.

\bibitem{Tsuji2016}
	N. Tsuji, Y. Murakami, and H. Aoki, ``Nonlinear light-Higgs coupling in superconductors beyond BCS: Effects of the retarded phonon-mediated interaction,'' Phys. Rev. B \textbf{94}, 224519 (2016).

\bibitem{Chou2017}
	Y.-Z. Chou, Y. Liao, and M. S. Foster, ``Twisting Anderson pseudospins with light: Quench dynamics in terahertz-pumped BCS superconductors,'' Phys. Rev. B \textbf{95}, 104507 (2017).

\bibitem{Kennes2017}
	D. M. Kennes, E. Y. Wilner, D. R. Reichman, and A. J. Millis, ``Nonequilibrium optical conductivity: General theory and application to transient phases,'' Phys. Rev. B \textbf{96}, 054506 (2017).

\bibitem{Schrieffer1964}
	J. R. Schrieffer, \textit{Theory of Superconductivity} (Benjamin, New York, 1964).

\bibitem{BCS1957}
	J. Bardeen, L. N. Cooper, and J. R. Schrieffer, ``Theory of superconductivity,'' Phys. Rev. \textbf{108}, 1175 (1957).

\bibitem{Coleman2015}
	P. Coleman, ``Introduction to Many-Body Physics'' (Cambridge University Press, 2015).

\bibitem{Klein1984}
	M. V. Klein and S. B. Dierker, ``Theory of Raman scattering in superconductors,'' Phys. Rev. B \textbf{29}, 4976 (1984).

\bibitem{Murakami2016}
	Y. Murakami, P. Werner, N. Tsuji, and H. Aoki, ``Damping of the collective amplitude mode in superconductors with strong electron-phonon coupling,'' Phys. Rev. B \textbf{94}, 115126 (2016).

\bibitem{Moor2017}
	A. Moor, A. F. Volkov, and K. B. Efetov, ``Amplitude Higgs mode and admittance in superconductors with a moving condensate,'' Phys. Rev. Lett. \textbf{118}, 047001 (2017).

\bibitem{Nakamura2018}
	S. Nakamura, Y. Iida, Y. Murotani, R. Matsunaga, H. Terai, and R. Shimano, ``Infrared activation of Higgs mode by supercurrent injection in a superconductor NbN,'' arXiv: 1809.10335.

\bibitem{Sung1967}
	C. C. Sung and V. K. Wong, ``Influence of nonmagnetic impurities on superconductors with overlapping bands,'' J. Phys. Chem. Solids \textbf{28}, 1933 (1967).

\end{thebibliography}
\end{document}